\documentclass[11pt]{report}
\pdfoutput=1

\usepackage{a4wide}

\usepackage{Packages}

\usepackage{Definitions}
\usepackage{multirow}
\usepackage{comment}

\usepackage{minitoc}
\usepackage[sectionbib]{chapterbib}

\hypersetup{pdfauthor={S.Reffert}, pdftitle={The Geometer's Toolkit to String Compactifications}, pdfkeywords={string compactifications, Calabi-Yau, toroidal orbifolds, toric geometry, resolved toroidal orbifolds, orientifold}}

\begin{document}
\begin{titlepage}

 \vspace{-15cm}
  \begin{flushright}
    ITFA-2007-24
  \end{flushright}

\vskip1.5cm
\begin{center}
  \textsc{\bf \Huge The Geometer's Toolkit\\ 
    to \\[2mm]
    String Compactifications}
  
  \vskip2.5cm

  \vskip 0.2cm 
  based on lectures given at the\\[18pt]
  { \large Workshop on \\[10pt]
    String and M--Theory Approaches to\\[10pt]
    Particle Physics and Astronomy\\[25pt]
    Galileo Galilei Institute for Theoretical Physics\\[10pt] 
    Arcetri (Firenze)}

  \vskip 5cm 
  {\large Susanne Reffert\\[6pt]
    ITFA Amsterdam}
  
  \vskip 2cm
  {\large May 2007}
\end{center}

\end{titlepage}
\newpage

\dominitoc
\tableofcontents


\chapter*{Summary}
\label{sec:summary}


The working string theorist is often confronted with the need to make
use of various techniques of algebraic geometry. Unfortunately, a
problem of language exists. The specialized mathematical literature is
often difficult to read for the physicist, moreover differences in
terminology exist.

These lectures are meant to serve as an introduction to some geometric constructions and techniques (in particular the ones of toric geometry) often employed by the physicist working on string theory compactifications. The emphasis is wholly on the geometry side, not on the physics. Knowledge of the basic concepts of differential, complex and K\"ahler geometry is assumed.

The lectures are divided into four parts. Lecture one briefly reviews the basics of Calabi--Yau geometry and then introduces toroidal orbifolds, which enjoy a lot of popularity in string model building constructions.
In lecture two, the techniques of toric geometry are introduced, which are of vital importance for a large number of Calabi--Yau constructions. In particular, it is shown how to resolve orbifold singularities, how to calculate the intersection numbers and how to determine divisor topologies.
In lectures three, the above techniques are used to construct a smooth Calabi--Yau manifold from toroidal orbifolds by resolving the singularities locally and gluing together the smooth patches. 
The full intersection ring and the divisor topologies are determined by a combination of knowledge about the global structure of $T^6/\Gamma$ and toric techniques. In lecture four, the orientifold quotient of such a resolved toroidal orbifold is discussed.

The theoretical discussion of each technique is followed by a simple, explicit example.
At the end of each lecture, I give some useful references, with emphasis on text books and review articles, not on the original articles.

\chapter{Calabi--Yau basics and orbifolds}
\minitoc
\bigskip
In this lecture, I will briefly review the basics of Calabi--Yau geometry. As a simple and extremely common example, I will introduce toroidal orbifolds.

\section{Calabi--Yau manifolds}

Calabi conjectured in 1957 that a compact K\"ahler manifold $X$ of vanishing first Chern class always admits  a Ricci--flat metric. This was proven by Yau in 1977. Such a manifold $X$ of dimension $n$ is now known as \emph{Calabi--Yau} manifold. 
Equivalently, $X$ is Calabi--Yau if it
\begin{itemize}
\item[(a)] admits a Levi--Civita connection with $SU(n)$ holonomy
\item[(b)] admits a nowhere vanishing holomorphic $(n,0)$--form $\Omega$
\item[(c)] has a trivial canonical bundle.
\end{itemize}

The Hodge numbers of a complex manifold are often displayed in a so--called \emph{Hodge diamond}:
\begin{equation}
  \begin{tabular}{p{.3em}p{.3em}p{.3em}p{.3em}p{.3em}p{.3em}p{.3em}}
    & & & $h^{0,0}$ \\
    & & $h^{1,0}$ & & $h^{0,1}$ \\
    & $h^{2,0}$ & & $h^{1,1}$ & & $h^{0,2}$ \\
    $h^{3,0}$ & & $h^{2,1}$ & & $h^{1,2}$ & & $h^{0,3}$ \\
    & $h^{3,1}$ & & $h^{2,2}$ & & $h^{1,3}$ \\
    & & $h^{3,2}$ & & $h^{2,3}$ \\
    & & & $h^{3,3}$
  \end{tabular}
\end{equation}
For a K\"ahler manifold, the Hodge diamond has two symmetries:
\begin{itemize}
\item Complex conjugation $\Rightarrow\, h^{p,q}=h^{q,p}$ (vertical reflection symmetry),
\item Poincar\'e duality $\Rightarrow\, h^{p,q}=h^{n-q,n-p}$
  (horizontal reflection symmetry).
\end{itemize} 
For $X$ being Calabi--Yau, the Hodge diamond is even more constrained:
(b) implies that $h^{n,0}=1$ and furthermore $h^{p,0}=h^{n-p,0}$.  The
Hodge--diamond of a Calabi--Yau 3--fold therefore takes the form
\begin{equation}
  \begin{tabular}{p{.3em}p{.3em}p{.3em}p{.3em}p{.3em}p{.3em}p{.3em}}
    & & & $1$ \\
    & & $0$ & & $0$ \\
    & $0$ & & $h^{1,1}$ & & $0$ \\
    $1$ & & $h^{2,1}$ & & $h^{2,1}$ & & $1$ \\
    & $0$ & & $h^{1,1}$ & & $0$ \\
    & & $0$ & & $0$ \\
    & & & $1$
  \end{tabular}
\end{equation}
Thus, the Hodge numbers of $X$ are completely specified by $h^{1,1}$
and $h^{2,1}$. The Euler number of $X$ is
\begin{equation}
  \chi(X)=2\,(h^{1,1}(X)-h^{2,1}(X)).
\end{equation}
Until fairly recently, not a single example of an explicit compact
Calabi--Yau metric was known! \footnote{This only changed with the
  introduction of Calabi--Yaus that are cones over Sasaki--Einstein
  manifolds, see \eg\ \cite{Gauntlett:2004yd}.}

A Calabi--Yau manifold can be deformed in two ways: Either by varying
its \emph{complex structure} (its "shape"), or by varying its
\emph{K\"ahler structure} (its "size"). Variations of the metric of
mixed type $\delta g_{m \ov n}$ correspond to variations of the
K\"ahler structure and give rise to $h_{1,1}$ parameters, whereas
variations of pure type $\delta g_{mn},\,\delta g_{\ov m\ov n}$
correspond to variations of the complex structure and give rise to
$h_{2,1}$ complex parameters.  To metric variations of mixed type, a
real $(1,1)$--form can be associated:
\begin{equation}
  i\,\delta g_{m\ov n}\,dz^m\wedge d\ov z^{\ov n} \, .
\end{equation}
To pure type metric variations, a complex $(2,1)$--form can be
associated:
\begin{equation}
  \Omega_{ijk}\,g^{k\ov n}\,\delta g_{\ov m\ov n}\,dz^i\wedge dz^j\wedge d\ov z^{\ov m} \, ,
\end{equation}
where $\Omega$ is the Calabi--Yau $(3,0)$--form.

\subsubsection{1D Calabi--Yaus}

It is easy to list all one--dimensional Calabi--Yaus: there is but the complex plane, the punctured complex plane (\ie the cylinder) and the two--torus $T^2$.

The Hodge diamond of a 1D Calabi--Yau is (not surprisingly) completely constrained:
\begin{equation}
  \begin{tabular}[c]{ccc}
    \begin{minipage}{6em}
      \begin{center}
        \begin{tabular}{p{.3em}p{.3em}p{.3em}}
          & $h^{0,0}$ \\
          $h^{1,0}$ &  & $h^{0,1}$ \\
          & $h^{1,1}$ 
        \end{tabular}
      \end{center}
    \end{minipage} & = & 
    \begin{minipage}{4em}
      \begin{center}
        \begin{tabular}{p{.3em}p{.3em}p{.3em}}
          & $1$ \\
          $1$ &  & $1$ \\
          & $1$ 
        \end{tabular}
      \end{center}
    \end{minipage}
  \end{tabular}  
\end{equation}
We now illustrate the concept of moduli for the simple case of $T^2$,
which has the metric
\begin{equation}\label{mm}{
    g=\left(\begin{array}{cc}
        g_{11}&g_{12}\cr g_{12}&g_{22}\end{array}\right)=\left(\begin{array}{cc}
        R_1^2&R_1R_2\,\cos\theta_{12}\cr R_1R_2\,\cos\theta_{12}&R_2^2\end{array}\right).}
\end{equation}
A $T^2$ comes with one K\"ahler modulus $\mathcal{ T}$, which parametrizes its volume, and one complex structure modulus, which corresponds to its modular parameter $\mathcal{ U}=\tau$.
\begin{figure}[h!]
\begin{center}
\includegraphics[width=.4\textwidth]{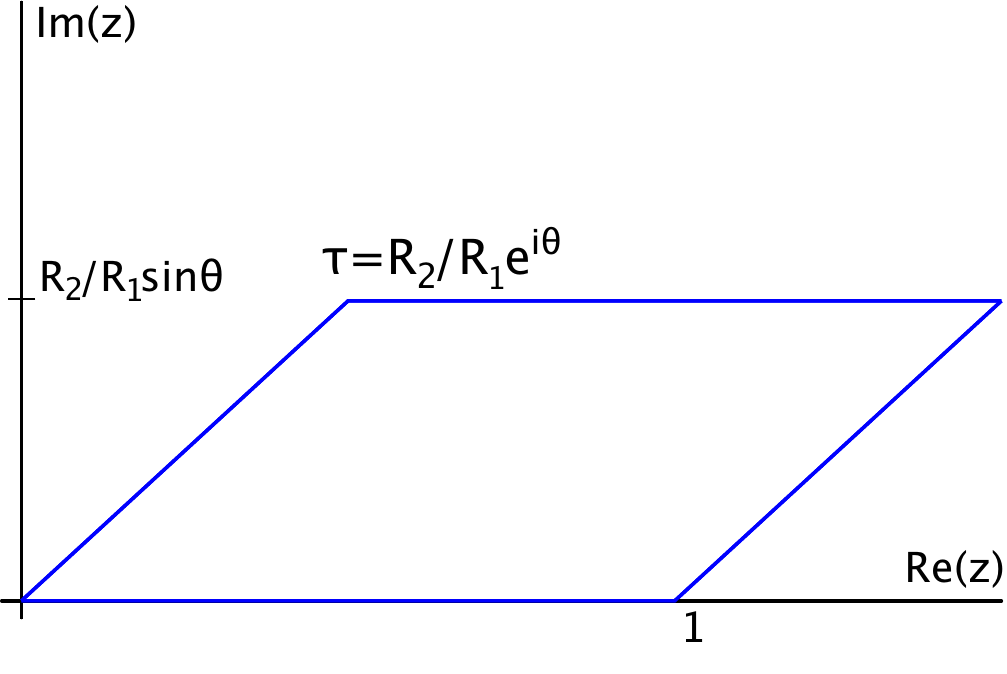}
\caption{Fundamental region of a $T^2$}
\label{fundamental}
\end{center}
\end{figure}
Figure \ref{fundamental} depicts the fundamental region of a $T^2$. The area of the torus is given by $R_1R_2\,\sin\theta_{12}$, expressed through the metric, we find 
\begin{equation}
\mathcal{ T}=\sqrt{\det\, g}=R_1R_2\,\sin\theta_{12}.
\end{equation}
In heterotic string theory, the K\"ahler moduli are complexified by pairing them up with the components of the anti--symmetric tensor $B$.
In type $IIB$ string theory, the K\"ahler moduli are paired with the components of the Ramond--Ramond four--form $C_4$.
The usual normalization of the fundamental region in string theory is such that the $a$--cycle is normalized to 1, while the modular parameter becomes $\tau=R_2/R_1\,e^{i\theta}$.
The complex structure modulus expressed through the metric is
\begin{equation}\label{CS}
\mathcal{U}=\frac{1}{{g_{11}}}\ (\,g_{12}+i\,\sqrt{\det\, g}\,).
\end{equation}

\subsubsection{2D Calabi--Yaus}

In two dimensions, there are (up to diffeomorphism) only two compact Calabi--Yaus: the $K3$--surface and the 4--torus $T^4$.
The Hodge diamond of the $K3$ is
\bigskip
\begin{equation}
  \begin{tabular}[c]{ccc}
    \begin{minipage}{8.5em}
      \begin{center}
        \begin{tabular}{p{.3em}p{.3em}p{.3em}p{.3em}p{.3em}}
          & & $h^{0,0}$ \\
          &  $h^{1,0}$ &  & $h^{0,1}$ \\
          $h^{2,0}$ & &  $h^{1,1}$ &  & $h^{0,2}$\\
          &  $h^{2,1}$ &  & $h^{1,2}$ \\
          &  &$h^{2,2}$ 
        \end{tabular}
      \end{center}
    \end{minipage} & = & 
    \begin{minipage}{6em}
      \begin{center}
        \begin{tabular}{p{.3em}p{.3em}p{.3em}p{.3em}p{.3em}}
          & & $1$ \\
          & $0$ &  & $0$ \\
          $1$ &   & $20$ &  & $1$\\
          & $0$ &  & $0$ \\
          & &$1$ 
        \end{tabular}
      \end{center}
    \end{minipage}
  \end{tabular}  
\end{equation}

\subsubsection{3D Calabi--Yaus}

In three dimensions, no classification exists. It is not even known
whether there are finitely or infinitely many (up to diffeomorphism).

There are several classes, which can be constructed fairly easily:
\begin{itemize}
\item hypersurfaces in toric varieties
\item complete intersections in toric varieties (CICY)
\item toroidal orbifolds and their resolutions
\item Cones over Sasaki--Einstein spaces (with metric!).
\end{itemize}

In the following, I will mainly concentrate on the third point. The
machinery of toric geometry will be introduced, which is vital for
most of these constructions.

\section{Orbifolds -- A simple and common example}

The string theorist has been interested in orbifolds for many years already (see \cite{Dixon:1985jw} in 1985), and for varying reasons, one of course being their simplicity. Knowledge of this construction is thus one of the basic requirements for a string theorist.

An \emph{orbifold} is obtained by dividing a smooth manifold by the
non--free action of a discrete group:
\begin{equation}
  X=Y/\Gamma \, .
\end{equation}
The original mathematical definition is broader: any algebraic variety
whose only singularities are locally of the form of quotient
singularities is taken to be an orbifold.

The string theorist is mostly concerned with \emph{toroidal orbifolds}
of the form $T^6/\Gamma$. While the torus is completely flat, the
orbifold is flat almost everywhere: its curvature is concentrated in
the fixed points of $\Gamma$. At these points, conical singularities
appear.

Only the simplest variety of toroidal orbifolds will be discussed here: $\Gamma$ is taken to be \emph{abelian}, there will be \emph{no discrete torsion or vector structure}.

Toroidal orbifolds are simple, yet non--trivial. Their main asset is calculability, which holds for purely geometric as well as for string theoretic aspects.

\subsection{Point groups and Coxeter elements}

A \emph{torus} is specified by its underlying lattice $\Lambda$: Points which differ by a lattice vector are identified:
\begin{equation}
  x \sim x+l,\quad l\in \Lambda \, .
\end{equation}
The six--torus is therefore defined as quotient of $\IR^6$ with respect to the lattice $\Lambda$:
\begin{equation}
  T^6=\IR^6/\Lambda \, .
\end{equation}
To define an orbifold of the torus, we divide by a discrete group
$\Gamma$, which is called the \emph{point group}, or simply the
orbifold group. We cannot choose any random group as the point group
$\Gamma$, it must be an automorphism of the torus lattice $\Lambda$,
\ie\ it must preserve the scalar product and fulfill
\begin{equation}
  g\,l\in \Lambda\ \mathrm{if}\ l\in \Lambda,\ g\in\Gamma \, .
\end{equation}
To fully specify a toroidal orbifold, one must therefore specify both
the torus lattice as well as the point group. In the context of string
theory, a set--up with $SU(3)$--holonomy\footnote{This results in
  $\mathcal{ N}=1$ supersymmetry for heterotic string theory and in
  $\mathcal{ N}=2$ in type $II$ string theories in four dimensions.}
is what is usually called for, which restricts the point group
$\Gamma$ to be a subgroup of $SU(3)$. Since we restrict ourselves to
abelian point groups, $\Gamma$ must belong to the Cartan subalgebra of
$SO(6)$. On the complex coordinates of the torus, the orbifold twist
will act as
\begin{equation}\label{cplxtwist}
\theta:\ (z^1,\,z^2,\,z^3) \to (e^{2\pi i \zeta_1}\,z^1,\, e^{2\pi i  \zeta_2}\,z^2,\, e^{2\pi i  \zeta_3}\,z^3), \quad 0\leq| \zeta^i|<1,\ i=1,2,3.
\end{equation}
The requirement of $SU(3)$--holonomy can also be phrased as requiring invariance of the $(3,0)$--form of the torus, $\Omega=dz^1\wedge dz^2\wedge dz^3$.
This leads to
\begin{equation}\label{req}
\pm  \zeta_1\pm  \zeta_2\pm  \zeta_3=0.
\end{equation}
We must furthermore require that $\Gamma$ acts crystallographically on the torus lattice. Together with the condition (\ref{req}), this amounts to $\Gamma$ being
either 
\begin{equation}\label{eq:zn}
  \IZ_N \text{ with } N=3,4,6,7,8,12 \, ,
\end{equation}
or $\IZ_N\times \IZ_M$ with $M$ a multiple of $N$ and
$N=2,3,4,6$. With the above, one is lead to the usual standard
embeddings of the orbifold twists, which are given in Tables
\ref{table:std} and \ref{table:zzstd}. The most convenient notation is
\begin{equation}
  ( \zeta_1, \zeta_2, \zeta_3)=\frac{1}{n}(n_1,n_2,n_3)\ \mathrm{ with}\ n_1+n_2+n_3=0\,\mod\, n \, . 
\end{equation}
Notice that $\IZ_6,\ \IZ_8$ and $\IZ_{12}$ have two inequivalent
embeddings in $SO(6)$.

\begin{table}[h!]\begin{center}
\begin{tabular}{lc}
\toprule
Point group & $\frac{1}{n}(n_1,n_2,n_3)$ \\ [2pt]
\midrule
 \rowcolor[gray]{.95}$\ \IZ_{3}  $      & $\frac{1}{3}\,(1,1,-2)$\\ [2pt]
$\ \IZ_{4}  $      & $\frac{1}{4}\,(1,1,-2)$\\ [2pt]
 \rowcolor[gray]{.95}$\ \IZ_{6-I}  $      & $\frac{1}{6}\,(1,1,-2)$\\ [2pt]
$\ \IZ_{6-II}  $      & $\frac{1}{6}\,(1,2,-3)$\\ [2pt]
 \rowcolor[gray]{.95}$\ \IZ_{7}  $      & $\frac{1}{7}\,(1,2,-3)$\\ [2pt]
$\ \IZ_{8-I}  $      & $\frac{1}{8}\,(1,2,-3)$\\ [2pt]
 \rowcolor[gray]{.95}$\ \IZ_{8-II}  $      & $\frac{1}{8}\,(1,3,-4)$\\ [2pt]
$\ \IZ_{12-I}  $      & $\frac{1}{12}\,(1,4,-5)$\\ [2pt]
 \rowcolor[gray]{.95}$\ \IZ_{12-II}  $      & $\frac{1}{12}\,(1,5,-6)$\\ [2pt]
\bottomrule
\end{tabular}
\caption{Group generators for $\IZ_N$-orbifolds.}\label{table:std}
\end{center}\end{table}

\begin{table}[h!]\begin{center}
\begin{tabular}{lcc}
\toprule
Point group & $\frac{1}{n}(n_1,n_2,n_3)$ & $\frac{1}{m}(m_1,m_2,m_3)$ \\ [2pt]
\midrule
 \rowcolor[gray]{.95}$\ \IZ_{2} \times \IZ_{2} $      & $\frac{1}{2}\,(1,0,-1)$ & $\frac{1}{2}\,(0,1,-1)$\\ [2pt]
$\ \IZ_{2} \times \IZ_{4} $      & $\frac{1}{2}\,(1,0,-1)$ & $\frac{1}{4}\,(0,1,-1)$\\ [2pt]
 \rowcolor[gray]{.95}$\ \IZ_{2} \times \IZ_{6} $      & $\frac{1}{2}\,(1,0,-1)$ & $\frac{1}{6}\,(0,1,-1)$\\ [2pt]
$\ \IZ_{2} \times \IZ_{6'} $      & $\frac{1}{2}\,(1,0,-1)$ & $\frac{1}{6}\,(1,1,-2)$\\ [2pt]
 \rowcolor[gray]{.95}$\ \IZ_{3} \times \IZ_{3} $      & $\frac{1}{3}\,(1,0,-1)$ & $\frac{1}{3}\,(0,1,-1)$\\ [2pt]
$\ \IZ_{3} \times \IZ_{6} $      & $\frac{1}{3}\,(1,0,-1)$ & $\frac{1}{6}\,(0,1,-1)$\\ [2pt]
 \rowcolor[gray]{.95}$\ \IZ_{4} \times \IZ_{4} $      & $\frac{1}{4}\,(1,0,-1)$ & $\frac{1}{4}\,(0,1,-1)$\\ [2pt]
$\ \IZ_{6} \times \IZ_{6} $      & $\frac{1}{6}\,(1,0,-1)$ & $\frac{1}{6}\,(0,1,-1)$\\ [2pt]
\bottomrule
\end{tabular}
\caption{Group generators for $\IZ_N\times \IZ_M$-orbifolds.}\label{table:zzstd}
\end{center}\end{table}

For all point groups given in Tables \ref{table:std} and
\ref{table:zzstd}, it is possible to find a compatible torus lattice,
in several cases even more than one. 

We will now repeat the same construction starting out from a real six--dimensional lattice. A lattice is suitable for our purpose if its automorphism group contains subgroups in $SU(3)$. Taking the eigenvalues of the resulting twist, we are led back to twists of the form (\ref{eq:zn}). A possible choice is to consider the
\emph{root lattices of semi--simple Lie--Algebras of rank 6}.  All one
needs to know about such a lattice is contained in the \emph{Cartan
  matrix} of the respective Lie algebra. The matrix elements of the
Cartan matrix are defined as follows:
\begin{equation}
  A_{ij}=2\ {\langle e_i, e_j\rangle\over \langle e_j, e_j\rangle} \, ,
\end{equation}
where the $e_i$ are the simple roots. 

The inner automorphisms of these root lattices are given by the Weyl--group of the Lie--algebra.  A \emph{Weyl reflection} is a reflection on the hyperplane perpendicular to a given root:
\begin{equation}\label{Weyl}
{S_i(\mathbf{ x})= \mathbf{ x}-2\frac{\langle\mathbf{ x}, e_i\rangle}{ \langle e_i,e_i\rangle}e_i}.
\end{equation}
These reflections are not in $SU(3)$ and therefore not suitable candidates for a point group, but the Weyl group does have a subgroup contained in $SU(3)$: the cyclic subgroup generated by the \emph{Coxeter element}, which is given by successive Weyl reflections with respect to all 
simple roots: 
\begin{equation}\label{cox}
Q=S_1S_2...S_{rank}.
\end{equation}
The so--called outer automorphisms are those which are generated by transpositions of roots which are symmetries of the Dynkin diagram. By combining Weyl reflections with such outer automorphisms, we arrive at so--called \emph{generalized Coxeter elements}.  $P_{ij}$ denotes the transposition of the $i$'th and $j$'th roots.
 
The orbifold twist $\Gamma$ may be represented by a matrix $Q_{ij}$,
which rotates the six lattice basis vectors:\footnote{Different
  symbols for the orbifold twist are used according to whether we look
  at the quantity which acts on the real six-dimensional lattice $(Q)$
  or on the complex coordinates $(\theta)$.}
\begin{equation}
  e_i\to Q_{ji}\ e_j \, .
\end{equation}
The following discussion is restricted to cases in which the orbifold
twist acts as the (generalized) Coxeter element of the group lattices,
these are the so--called \emph{ Coxeter--orbifolds}\footnote{It is
  also possible to construct non--Coxeter orbifolds, such as \eg\
  $\IZ_4$ on $SO(4)^3$ as discussed in \cite{Casas:1991ac}.}.

We now change back to the complex basis $\{z^i\}_{i=1,2,3}$, where the twist
$Q$ acts diagonally on the complex coordinates, \ie
\begin{equation}
  \theta\ :\ z^i\to e^{2\pi i \zeta_i} z^i \, ,\label{eq:28}
\end{equation}
with the eigenvalues $2\pi i\, \zeta_i$ introduced above.  To find these
complex coordinates we make the ansatz
\begin{equation}
  \label{ansatz}
  {z^i=a^i_1\,x^1+a^i_2\,x^2+a^i_3\,x^3+a^i_4\,x^4+a^i_5\,x^5+a^i_6\,x^6} \, .
\end{equation}
Knowing how the Coxeter twist acts on the root lattice and therefore
on the real coordinates $x^i$, and knowing how the orbifold twist acts
on the complex coordinates, see Tables \ref{table:one} and
\ref{table:two}, we can determine the coefficients $a_j^i$ by solving
\begin{equation}\label{solveeq}
Q^t\, z^i=e^{2\pi i \zeta_i}\,z^i.
\end{equation}
The transformation which takes us from the real to the complex basis must be unimodular. The above equation only constrains the coefficients up to an overall complex normalization factor. For convenience we choose a normalization such that the first term is real.

\begin{exa}
  We take the torus lattice to be the root lattice of $G_2^2\times
  SU(3)$, a direct product of three rank two root lattices, and
  explicitly construct its Coxeter element.  First, we look at the
  $SU(3)$--factor. With the Cartan matrix of $SU(3)$,
  \begin{equation}
    A=\left(\begin{array}{cc}2&\!\!\!-1\\ \!\!\!-1&2\end{array}\right),
  \end{equation}
  and eq.~(\ref{Weyl}), the matrices of the two Weyl reflections can
  be constructed:
  \begin{equation}
    S_1=\left(\begin{array}{cc}\!\!-1&1\\ 0&1\end{array}\right),\quad S_2=\left(\begin{array}{cc}1&0\\ 1&\!\!-1\end{array}\right).
  \end{equation}
  The Coxeter element is obtained by multiplying the two:
  \begin{equation}
    Q_{SU(3)}=S_1S_2=\left(\begin{array}{cc}0&-1\\ 1&-1\end{array}\right).
  \end{equation}
  In the same way, we arrive at the Coxeter-element of $G_2$. The
  six-dimensional Coxeter element is built out of the three $2\times
  2$--blocks:
  \begin{equation}\label{Qsixi}
    Q= \begin{pmatrix}
        2&-1&0&0&0&0\\ 
        3&-1&0&0&0&0\\
        0&0&2&-1&0&0\\
        0&0&3&-1&0&0\\
        0&0&0&0&0&-1\\
        0&0&0&0&1&-1\end{pmatrix} \, .
  \end{equation}
  The eigenvalues of $Q$ are $e^{2\pi i/6},\, e^{-2\pi i/6},\,e^{2\pi
    i/6},\, e^{-2\pi i/6},\,e^{2\pi i/3},\, e^{-2\pi i/3}$, \ie\ those
  of the $\IZ_{6-I}$--twist, see Table \ref{table:std}, and $Q$
  fulfills $Q^6=\mathrm{ Id}$.

  Solving (\ref{solveeq}) yields the following solution for the
  complex coordinates:
  \begin{align}
    z^1 &= a\,(-(1+e^{2\pi i/6})\,x^1+x^2)+b\,(-(1+e^{2\pi i/6})\,x^3+x^4),\cr
    z^2 &= c\,(-(1+e^{2\pi i/6})\,x^1+x^2)+d\,(-(1+e^{2\pi i/6})\,x^3+x^4),\cr
    z^3 &= e\,(e^{2\pi i/3}\,x^5+x^6),
  \end{align}
  where $a,\,b,\,c,\,d$ and $e$ are complex constants left unfixed by
  the twist alone. In the following, we will choose $a,\,d,\,e$ such
  that $x^1,\,x^3,\,x^5$ have a real coefficient and the
  transformation matrix is unimodular and set $b=c=0$, so the complex
  structure takes the following form:
  \begin{align}\label{cplxzsixisim}
    z^1 &= x^1+{1\over \sqrt3}\,e^{5\pi i/6}\,x^2,\cr
    z^2 &= x^3+{1\over \sqrt3}\,e^{5\pi i/6}\,x^4,\cr
    z^3 &= 3^{1/4}\,(x^5+e^{2\pi i/3}\,x^6).
  \end{align}

\end{exa}

\subsection{List of point groups and lattices}

In the Tables \ref{table:one} and \ref{table:two}, a list of torus lattices together with the compatible orbifold point group is given \cite{Erler:1992ki}.\footnote{Other references such as \cite{Bailin:1999nk} give other lattices as well.}
Notice that some point groups are compatible with several lattices. 

\begin{table}[p]
\begin{center}
\begin{center}
{\small
\begin{tabular}{lcccccc}
\toprule
$\ \IZ_N$&Lattice &$h_{(1,1)}^\mathrm{ untw.}$&$h_{(2,1)}^\mathrm{ untw.}$&
$h_{(1,1)}^\mathrm{ twist.}$&$h_{(2,1)}^\mathrm{ twist.}$ \\ [2pt]
\midrule
 \rowcolor[gray]{.95}$\ \IZ_3 $    &\  $SU(3)^3 $          &9 & 0 & 27 & 0\cr
$\  \IZ_4  $   &\ $SU(4)^2  $      &5 & 1 & 20 & 0\cr
 \rowcolor[gray]{.95}$\  \IZ_4 $    &\  $SU(2)\times SU(4)\times SO(5)$ &5 & 1 & 22 & 2\cr
$\  \IZ_4$      &\  $SU(2)^2\times SO(5)^2 $ &5 & 1 & 26 & 6\cr
 \rowcolor[gray]{.95}$\  \IZ_{6-I} $   & $(G_2\times SU(3)^{2})^{\flat}$  &5 & 0 & 20 & 1\cr
$\  \IZ_{6-I}  $  &$ SU(3)\times G_2^2$ &5 & 0 & 24 & 5\cr
 \rowcolor[gray]{.95}$\  \IZ_{6-II} $  &$ SU(2)\times SU(6)$    &3 & 1 & 22 & 0\cr
$\  \IZ_{6-II} $  &$SU(3)\times SO(8) $&3 & 1 & 26 & 4\cr
 \rowcolor[gray]{.95}$\  \IZ_{6-II} $  &$(SU(2)^2\times SU(3)\times SU(3))^{\sharp} $ &3 & 1 & 28 & 6\cr
$\  \IZ_{6-II}  $ &$ SU(2)^2\times SU(3)\times G_2$  &3 & 1 & 32 & 10\cr
 \rowcolor[gray]{.95}$\  \IZ_7  $      &$ SU(7) $                             &3 & 0 & 21 & 0\cr
$\  \IZ_{8-I}  $  &$ (SU(4)\times SU(4))^*$                              &3 & 0 & 21 & 0\cr
 \rowcolor[gray]{.95}$\  \IZ_{8-I}  $  &$SO(5)\times SO(9)    $     &3 & 0 & 24 & 3\cr
$\  \IZ_{8-II} $  &$ SU(2)\times SO(10)  $     &3 & 1 & 24 & 2\cr
 \rowcolor[gray]{.95}$\  \IZ_{8-II}  $   &$ SO(4)\times SO(9)$   &3 & 1 & 28 & 6\cr
$\  \IZ_{12-I}$   &$ E_6 $  &3 & 0 & 22 & 1\cr
 \rowcolor[gray]{.95}$\  \IZ_{12-I} $  &$ SU(3)\times F_4$  &3 & 0 & 26 & 5\cr
$\  \IZ_{12-II} $  &$SO(4)\times F_4$  &3 & 1 & 28 & 6\cr
\bottomrule
\end{tabular}}
\end{center}
\caption{Twists, lattices and Hodge numbers for $\IZ_N$ orbifolds.}\label{table:one}
\end{center}\end{table}

The tables give the torus lattices and the twisted and untwisted Hodge numbers.
The lattices marked with $\flat$, $\sharp$, and $*$ are realized as generalized Coxeter twists, the 
automorphism being in the first and second case $S_1S_2S_3S_4P_{36}P_{45}$ and in the third  $S_1S_2S_3P_{16}P_{25}P_{34}$.

\begin{table}[p]\begin{center}
\begin{center}
\begin{tabular}{lcccccc}
\toprule
$\ \IZ_N $&Lattice&
 $h_{(1,1)}^\mathrm{ untw.} $& $h_{(2,1)}^\mathrm{ untw.} $& 
 $h_{(1,1)}^\mathrm{ twist.} $& $h_{(2,1)}^\mathrm{ twist.} $\\ [2pt]
 \midrule
 \rowcolor[gray]{.95}$\  \IZ_2 \times\IZ_2$     &$SU(2)^6$  &3 & 3 & 48 & 0\cr
$\  \IZ_2 \times\IZ_4$    &$SU(2)^2\times SO(5)^2$ &3 & 1& 58 & 0\cr
 \rowcolor[gray]{.95}$\  \IZ_2 \times\IZ_6 $    &$ SU(2)^2\times SU(3)\times G_2$&3 & 1 & 48 & 2\cr
$\  \IZ_2 \times\IZ_{6'}$&$ SU(3)\times G_2^2$&3 & 0 & 33 & 0\cr
 \rowcolor[gray]{.95}$\  \IZ_3 \times\IZ_3  $   &$SU(3)^3$   &3 & 0 & 81 & 0\cr
$\  \IZ_3 \times\IZ_6 $    &$SU(3)\times G_2^2 $&3 & 0 & 70 & 1\cr
 \rowcolor[gray]{.95}$\  \IZ_4 \times\IZ_4 $    &$SO(5)^3 $ &3 & 0 & 87 & 0\cr
$\  \IZ_6 \times\IZ_6  $   &$G_2^3$ &3 & 0 & 81 & 0\cr
\bottomrule 
\end{tabular}
\end{center}
\caption{Twists, lattices and Hodge numbers for $\IZ_N\times \IZ_M$ orbifolds.}\label{table:two}
\end{center}\end{table}

\subsection{Fixed set configurations and conjugacy classes}\label{sec:schematic}

Many of the defining properties of an orbifold are encoded in its singularities. 
Not only the type (which group element they come from, whether they are isolated or not) and number of singularities is important, but also their spatial configuration. Here, it makes a big difference on which torus lattice a specific twist lives. The difference does not arise for the fixed points in the first twisted sector, \ie\ those of the $\theta$--element which generates the group itself. But in the higher twisted sectors, in particular in those which give rise to fixed tori, the number of fixed sets differs for different lattices, which leads to differing Hodge numbers.

A point $f^{(n)}$ is fixed under $\theta^n\in \IZ_m,\ \ n=0,...,m-1,$ if it fulfills 
\begin{equation}\label{fix}{\theta^n\,f^{(n)}=f^{(n)}+l,\quad l\in \Lambda,
}\end{equation}
where $l$ is a vector of the torus lattice. In the real lattice basis, we have the identification 
\begin{equation}
  x^i \sim x^i + 1 \, .
\end{equation}
Like this, we obtain the sets that are fixed under the respective
element of the orbifold group.  A twist $\frac{1}{n}(n_1,n_2,n_3)$ and
its anti--twist $\frac{1}{n}(1-n_1,1-n_2,1-n_3)$ give rise to the same
fixed sets, so do permutations of $(n_1,n_2,n_3)$. Therefore not all
group elements of the point group need to be considered
separately. The prime orbifolds, \ie\ $\IZ_3$ and $\IZ_7$ have an
especially simple fixed point configuration since all twisted sectors
correspond to the same twist and so give rise to the same set of fixed
points.  Point groups containing subgroups generated by elements of
the form
\begin{equation}
  \frac{1}{n}\,(n_1, 0, n_2),\ n_1+n_2=0\,\mod\,n 
\end{equation}
give rise to \emph{fixed tori}.

It is important to bear in mind that the fixed points were determined on the covering space. On the quotient, points which form an orbit under the orbifold group are identified. For this reason, not the individual fixed sets, but their conjugacy classes must be counted.

To form a notion of what the orbifold looks like, it is useful to have a schematic picture of the configuration, \ie\ the intersection pattern of the singularities.


\begin{exa}
  In the following, we will identify the fixed sets under the
  $\theta$--, $\theta^2$-- and $\theta^3$--elements. $\theta^4$ and
  $\theta^5$ yield no new information, since they are simply the
  anti--twists of $\theta^2$ and $\theta$.  The $\IZ_{6-I}$--twist has
  only one fixed point in each torus, namely $z^i = 0$. The
  $\IZ_3$--twist has three fixed points in each direction, namely
  $z^1=z^2 = 0,1/3, 2/3$ and $z^3=0, 1/\sqrt3 \,e^{\pi i/6},
  1+i/\sqrt3$. The $\IZ_2$--twist, which arises in the
  $\theta^3$--twisted sector, has four fixed points, corresponding to
  $z^1=z^2 = 0,\half,\half \tau,\half(1+\tau)$ for the respective
  modular parameter $\tau$. As a general rule, we shall use red to
  denote the fixed set under $\theta$, blue to denote the fixed set
  under $\theta^2$ and pink to denote the fixed set under
  $\theta^3$. Note that the figure shows the covering space, not the
  quotient.

  Table~\ref{tab:fssixi} summarizes the important data of the fixed
  sets. The invariant subtorus under $\theta^3$ is $(0,0,0,0,x^5,x^6)$
  which corresponds simply to $z^3$ being invariant.

  \begin{table}[h!]
    \begin{center}
      \begin{tabular}{lccc}
        \toprule
        Group el.& Order &\ Fixed Set&Conj. Classes \cr
        \midrule
        \rowcolor[gray]{.95}$\ \theta=\tfrac{1}{6}(1,1,4)$&6&\ 3 fixed points &\ 3\\[.1cm]
        $\ \theta^2=\tfrac{1}{3}(1,1,1)$&3&\ 27 fixed points &\ 15\\[.1cm]
        \rowcolor[gray]{.95}$\ \theta^3=\tfrac{1}{2}(1,1,0)$&2&\ 16 fixed lines &\ 6\cr
        \bottomrule
      \end{tabular}
    \end{center}
    \caption{Fixed point set for $\IZ_{6-I}$ on $G_2^2\times SU(3)$}
    \label{tab:fssixi}
  \end{table}

  \begin{figure}[h!]
    \begin{center}
      \includegraphics[width=0.5\textwidth]{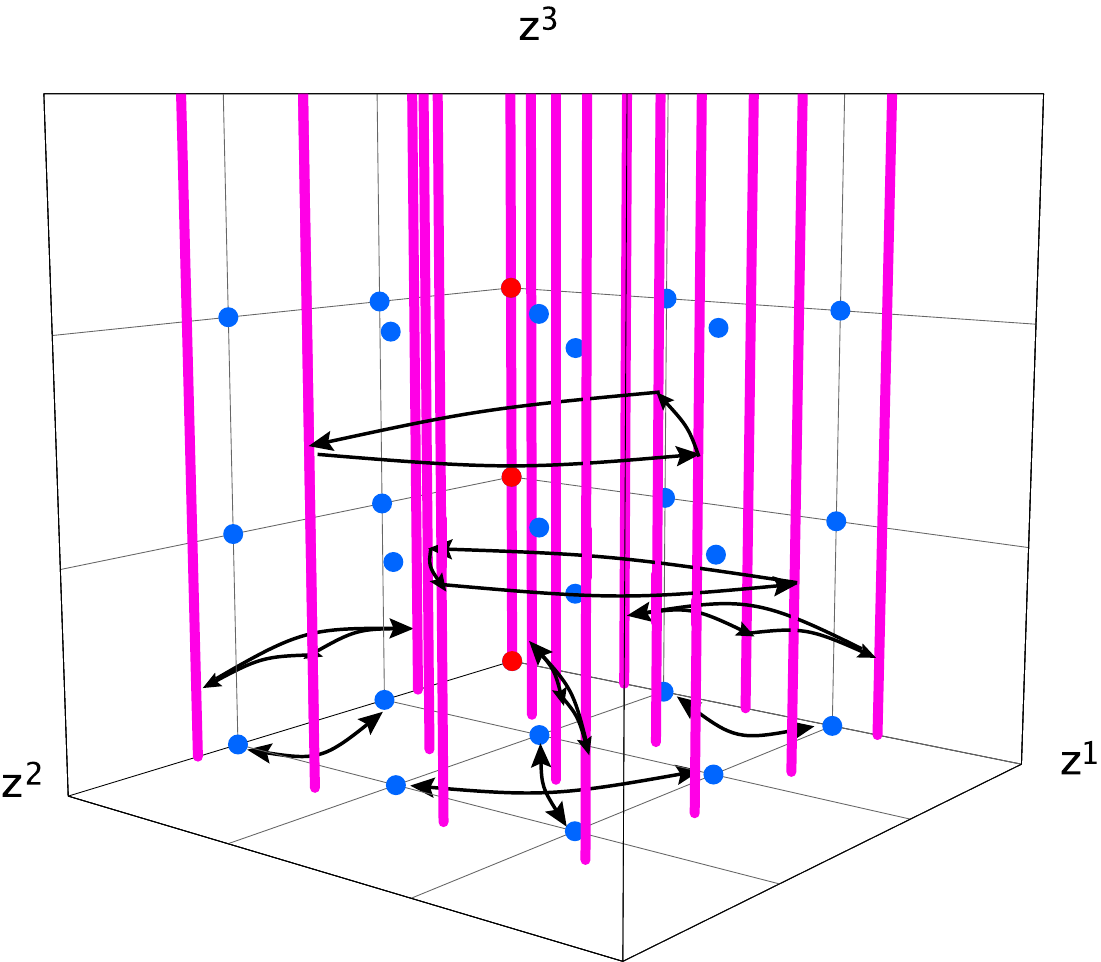}
      \caption{Schematic picture of the fixed set configuration of
        $\IZ_{6-I}$ on $G_2^2\times SU(3)$}
      \label{fig:ffixedi}
    \end{center}
  \end{figure}
  Figure \ref{fig:ffixedi} shows the configuration of the fixed sets
  in a schematic way, where each complex coordinate is shown as a
  coordinate axis and the opposite faces of the resulting cube of
  length 1 are identified. Note that this figure shows the whole
  six--torus and not the quotient.  The arrows indicate the orbits of
  the fixed sets under the action of the orbifold group.

\end{exa}

\section{Literature}

A very good introduction to complex manifolds are the lecture notes by
Candelas and de la Ossa \cite{candelas_lectures}, which are
unfortunately not available online. Usually, old paper copies which
were handed down from earlier generations can still be found in most
string theory groups.  For nearly all purposes, the book of Nakahara
\cite{Nakahara} is an excellent reference.  An introduction to all
necessary basics which is very readable for the physicist is given in
Part 1 of \cite{Hori}.  Specifically for Calabi--Yau geometry, there
is the book by H\"ubsch \cite{Huebsch}. A number of lecture notes and
reviews contain much of the basics, see for example
\cite{Greene:1996cy}.

On Orbifolds, a number of reviews exist (mainly focusing on physics,
though), \eg~\cite{Bailin:1999nk}. More orbifold examples as
introduced above are contained in \cite{Reffert:2006du}.

\bibliographystyle{JHEP}
\bibliography{LectureReferences}


\chapter{Toric Geometry}
\minitoc
\bigskip
In this lecture, I will introduce an extremely useful tool, namely the methods of \emph{toric geometry}. The geometry is summarized in combinatorial data, which is fairly simple to use.

After introducing the basics, I will discuss the resolution of singularities via a bow--up, the determination of the Mori generators and the intersection ring, as well as how to determine the divisor topologies in non--compact toric varieties. The material is introduced at the example of orbifolds of the form $\IC^3/\IZ_n$.

\section{The basics}

An $n$--dimensional \emph{toric variety} has the form
\begin{equation}\label{variety}{X_\Sigma=({\IC}^N\setminus F_\Sigma)/({\IC}^*)^m,}\end{equation}
where $m<N,\ \ n=N-m$. $({\IC}^*)^m$ is the \emph{algebraic torus} which lends the variety its name and acts via coordinatewise multiplication\footnote{An algebraic torus can be defined for any field $\mathbb{K}$. The name is connected to the fact that if  $\mathbb{K}=\IC$, an algebraic torus is the complexification of the standard torus $(S^1)^n$.}. $F_\Sigma$ is the subset that remains fixed under a continuous subgroup of $({\IC}^*)^m$ and must be subtracted for the variety to be well--defined. 

Toric varieties can also be described in terms of gauged linear sigma models. In short, for an appropriate choice of Fayet--Iliopoulos parameters, the space of supersymmetric ground states of the gauged linear sigma models is a toric variety. We will not take this point of view here and thus refer the reader to the literature, \eg~\cite{Hori}.

\begin{figure}[h!]
  \centering
  \begin{minipage}{.85\textwidth}
  \subfigure[Fan of $\IP^1$]{\includegraphics[width=.3\textwidth]{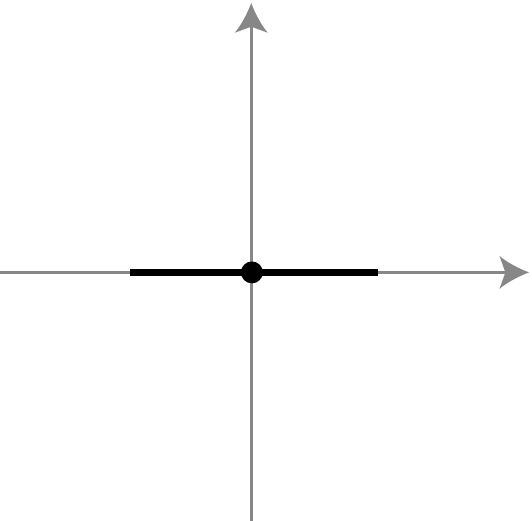}} \hfill
  \subfigure[Fan of $\IP^2$]{\includegraphics[width=.3\textwidth]{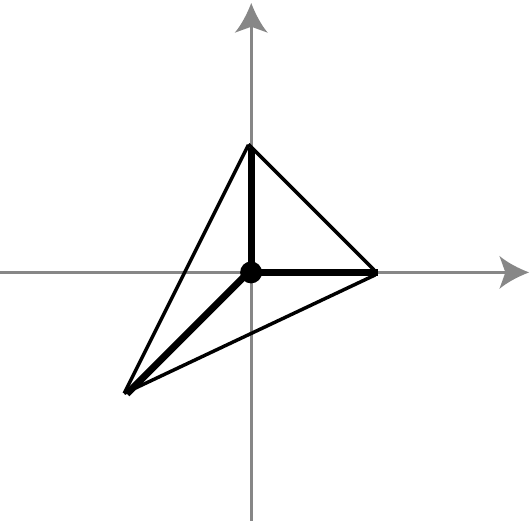}}   
  \end{minipage}
  \caption{Fans of projective spaces}
  \label{fig:projective}
\end{figure}

\newpage
\begin{ex0}
  The \emph{complex projective space} $\mathbb{P}^n$ (sometimes also
  denoted $\mathbb{CP}^n$) is defined by
  \begin{equation}
    \mathbb{P}^n=(\setC^{n+1}\setminus \{0\})/\setC^*.
  \end{equation}
  It is a quotient space and corresponds to the complex lines passing
  through the origin of $\setC^{n+1}$. $\setC^*$ acts by
  coordinatewise multiplication. 0 has to be removed, so $\setC^*$
  acts freely (without fixed points).  $\mathbb{P}^n$ thus corresponds
  to the space of $\setC^*$ orbits.  Points on the same line are
  equivalent:
  \begin{equation} [X_0, X_1, ..., X_n]\sim[\lambda\,X_0,\lambda\,
    X_1, ..., \lambda\,X_n], \quad \lambda\in\setC^*.
  \end{equation}
  The $X_0, ...X_n$ are the so--called \emph{homogeneous coordinates}
  and are redundant by one.  In a local coordinate patch with $X_i\neq
  0$, one can define coordinates invariant under rescaling
  \begin{equation}
    z_k=X_k/X_i, \quad k\neq i \, .
  \end{equation}
  $\IP^n$ is compact and all its complex submanifolds are
  compact. Moreover, Chow proved that any submanifold of $\IP^n$ can
  be realized as the zero locus of finitely many homogeneous
  polynomial equations. $\IP^1$ corresponds to $S^2$.

\emph{Weighted projective spaces} are a generalization of the above, with different torus actions:
\begin{equation}
\lambda:\, (X_0, X_1..., X_n)\mapsto (\lambda^{w_0}\, X_0,\, \lambda^{w_1}\,X_1, ...,\,\lambda^{w_n}\,X_n).
\end{equation}
With this $\lambda$ we can define
\begin{equation}
  \mathbb{P}^n_{(w_0,...,w_n)}= ( \setC^{n+1}\setminus\{0\} )/\setC^*.
\end{equation}
Note, that the action of $\setC^*$ is no longer free\footnote{Suppose $w_i\neq0$. Then it is possible to choose $\lambda\neq1$ such that $\lambda^{w_i}=1$, which results in $(0,..,0,X_i,0,...,0)=(0,...,0,\lambda^{w_i}X_i,0,...0)$.} . The weighted projective space will thus contain quotient singularities.

Projective spaces are obviously the most simple examples of toric varieties.
The fans (see Sec.~\ref{sec:fan}) of $\IP^1$ and $\IP^2$ are shown in Figure~\ref{fig:projective}.
\end{ex0}

\subsection*{A lattice and a fan}\label{sec:fan}

A toric variety $X_\Sigma$ can be encoded by a lattice $N$ which is isomorphic to $\IZ^n$ and its fan $\Sigma$. The \emph{fan} is a collection of strongly convex rational cones in $N\otimes_{\IZ}{\IR}$ with the property that each face of a cone in $\Sigma$ is also a cone in $\Sigma$ and the intersection of two cones in $\Sigma$ is a face of each. The $d$--dimensional cones in $\Sigma$ are in one--to--one correspondence with the codimension $d$--submanifolds of $X_\Sigma$. The one--dimensional cones in particular correspond to the divisors in $X_\Sigma$.

The fan $\Sigma$ can be encoded by the generators of its edges or one--dimensional cones, \ie\ by vectors $v_i\in N$. To each $v_i$ we associate a homogeneous coordinate $z_i$ of $X_\Sigma$.  To each of the $v_i$ corresponds the divisor $D_i$ which is determined by the equation $z_i=0$. The $(\IC^*)^m$ action on the $v_i$ is encoded  in $m$ linear relations
\begin{equation}
  \label{eq:linrels}
  \sum_{i=1}^d l^{(a)}_i v_i = 0, \qquad a=1,\dots,m, \quad l^{(a)}_i \in \IZ.
\end{equation}
To each linear relation we assign a monomial 
\begin{equation}
  U^a = \prod_{i=1}^d z_i^{l^{(a)}_i} \, .
\end{equation}
These monomials are invariant under the scaling action and form the local coordinates of $X_\sigma$. 
In general, monomials of type $z_1^{a_1}....z_k^{a_k}$ are sections of line bundles $\Oc(a_1\,D_1+...+a_k\,D_k)$. Let $M$ be the lattice dual to $N$ with respect to the pairing $\langle\ , \ \rangle$. For any $p\in M$, monomials of the form $z_1^{\langle v_1, p \rangle}....z_k^{\langle v_k, p \rangle}$ are invariant under the scaling action and thus give rise to a linear equivalence relation
\begin{equation}
\langle v_1, p \rangle\, D_1+...+\langle v_k, p \rangle\, D_k\sim 0\,.
\end{equation}

We are uniquely interested in Calabi--Yau manifolds, therefore we require $X_\Sigma$ to have trivial canonical class. The canonical divisor of  $X_\Sigma$ is given by $-D_1-...-D_n$, so for $X_\Sigma$ to be Calabi--Yau, $D_1+...+D_n$ must be trivial, \ie\ there must be a $p\in M$ such that  $\langle v_i, p \rangle=1$ for every $i$.
This translates to requiring that the $v_i$ must all lie in the same affine hyperplane one unit away from the origin $v_0$. In our 3--dimensional case, we can choose \eg\ the third component of all the vectors $v_i$ (except $v_0$) to equal one. The $v_i$ form a cone $C(\Delta^{(2)})$ over the triangle $\Delta^{(2)} = \langle v_1, v_2, v_3 \rangle$ with apex $v_0$. The Calabi--Yau condition therefore allows us to draw toric diagrams $\Delta^{(2)}$ in two dimensions only. The toric diagram drawn on the hyperplane has an obvious $SL(2,\IZ)$ symmetry, \ie\ toric diagrams which are connected by an $SL(2,\IZ)$ transformation give rise to the same toric variety.

In the \emph{dual diagram}, the geometry and intersection properties of a toric manifold are often easier to grasp than in the original toric diagram. The divisors, which are represented by vertices in the original toric diagram become faces in the dual diagram, the curves marking the intersections of two divisors remain curves and the intersections of three divisors which are represented by the faces of the original diagram become vertices. In the dual graph, it is immediately clear, which of the divisors and curves are compact.

For now, we remain with the orbifold examples discussed earlier. So how do we go about finding the fan of a specific ${\IC}^3/\IZ_n$--orbifold? We have just one three--dimensional cone in $\Sigma$, generated by $v_1,\,v_2,\,v_3$. The orbifold acts as follows on the coordinates of  ${\IC}^3$:
\begin{equation}\label{twistc}{\theta:\ (z_1,\, z_2,\, z_2) \to (\varepsilon\, z_1, \varepsilon^{n_1}\, z_2, \varepsilon^{n_2}\, z_3),\quad \varepsilon=e^{2 \pi i/n}.
}\end{equation}
For such an action we will use the shorthand notation $\frac{1}{n}(1,n_1,n_2)$. The coordinates of $X_\Sigma$ are given by
\begin{equation}\label{coord}{U^i=z_1^{(v_1)_i}z_2^{(v_2)_i}z_3^{(v_3)_i}.}
\end{equation}
To find the coordinates of the generators $v_i$ of the fan, we require the $U^i$ to be invariant under the action of $\theta$. We end up looking for two linearly independent solutions of the equation
\begin{equation}\label{eqfan}{(v_1)_i+n_1\,(v_2)_i+n_2\,(v_3)_i= 0\ \mod\, n.}
\end{equation}
The Calabi--Yau condition is trivially fulfilled since the orbifold actions are chosen such that $1+n_1+n_2=n$ and $\varepsilon^n=1$.

$X_\Sigma$ is smooth if all the top--dimensional cones in $\Sigma$ have volume one. By computing the determinant $\det(v_1,v_2,v_3)$, it can be easily checked that this is not the case in any of our orbifolds. We will therefore resolve the singularities by blowing them up.


\begin{exa1}
\label{sec:exzsixi}
  The group $\IZ_{6-I}$ acts as follows on $\IC^3$:
  \begin{equation}
    \label{eq:twistsixi}
    \theta:\ (z_1,\, z_2,\, z_3) \to (\varepsilon\, z_1, \varepsilon\, z_2, \varepsilon^4\, z_3),\quad \varepsilon=e^{2 \pi i/6}.
  \end{equation}
  To find the components of the $v_i$, we have to solve
  \begin{equation}
    (v_1)_i+(v_2)_i+4\,(v_3)_i=0\ \mod\,6 \, .
  \end{equation}
  This leads to the following three generators of the fan (or some
  other linear combination thereof):
  \begin{equation}
    v_1=\begin{pmatrix}1\\-2\\1\end{pmatrix},\ 
    v_2=\begin{pmatrix}-1\\-2\\1\end{pmatrix},\ 
    v_3=\begin{pmatrix}0\\1\\1\end{pmatrix}.
  \end{equation}
  The toric diagram of ${\IC}^3/\IZ_{6-I}$ and its dual diagram are
  depicted in Figure \ref{fig:sixib}.

\begin{figure}[h!]
  \begin{center}
    \includegraphics[width=0.7\textwidth]{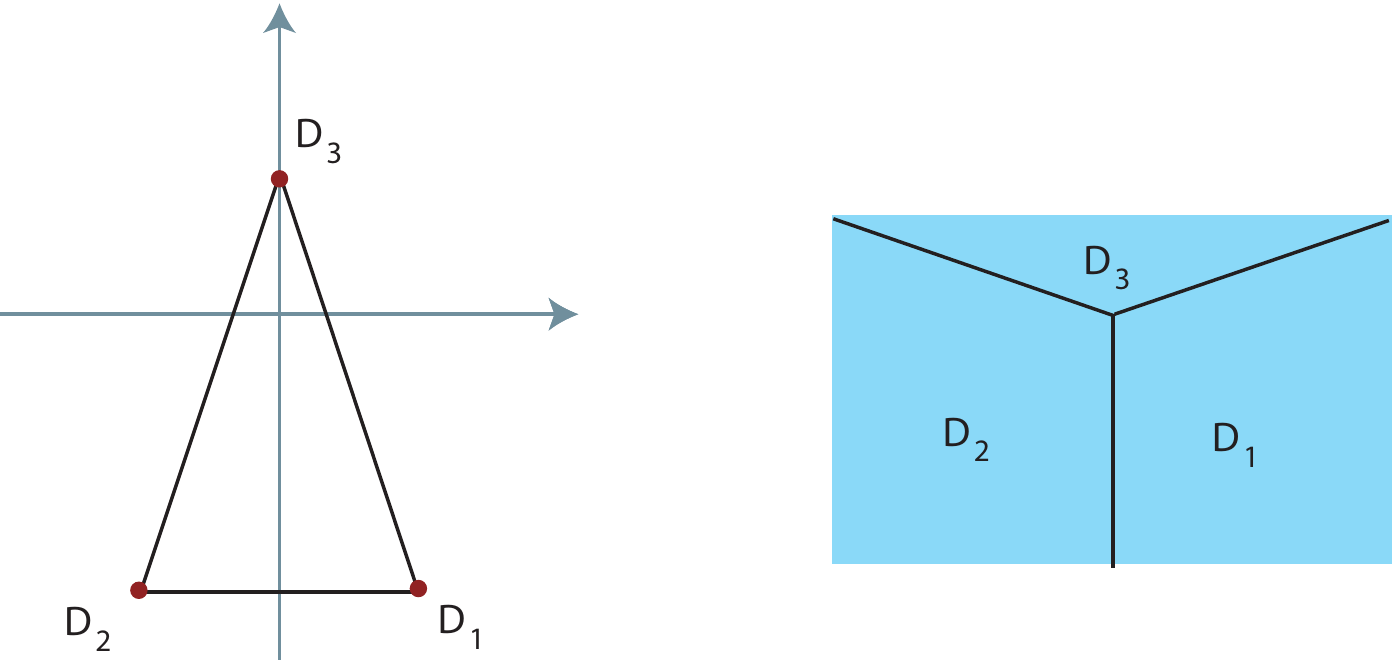}
    \caption{Toric diagram of $\IC^3/\IZ_{6-I}$ and dual graph}
    \label{fig:sixib}
  \end{center}
\end{figure}

\end{exa1}

\section{Resolution of singularities}\label{sec:resolve}

There are several ways of resolving a singularity, one of them being the blow--up.
The process of blowing up consists of two steps in toric geometry:
First, we must \emph{refine} the fan, then \emph{subdivide} it. Refining the fan means adding 1--dimensional cones. The subdivision corresponds to choosing a \emph{triangulation} for the toric diagram. Together, this corresponds to replacing the point that is blown up by an exceptional divisor. We denote the refined fan by $\tilde\Sigma$.

We are interested in resolving the orbifold--singularities such that the canonical class of the manifold is not affected, \ie\ the resulting manifold is still Calabi--Yau (in mathematics literature, this is called a \emph{crepant} resolution). 
When adding points that lie in the intersection of the simplex with corners $v_i$ and the lattice $N$, the Calabi--Yau criterion is met. Aspinwall studies the resolution of singularities of type ${\IC}^d/G$ and gives a very simple prescription \cite{Aspinwall:1994ev}. We first write it down for the case of  ${\IC}^3/\IZ_n$.
For what follows, it is more convenient to write the orbifold twists in the form
\begin{equation}{\theta:\ (z_1,\, z_2,\, z_3) \to (e^{2 \pi i g_1}\, z_1,\,e^{2 \pi i g_2}\, z_2,\,e^{2 \pi i g_3}\, z_3).
}\end{equation}
The new generators $w_i$ are obtained via
\begin{equation}\label{prescription}{
w_i=g^{(i)}_1\,v_1+g^{(i)}_2\,v_2+g^{(i)}_3\,v_3,
}\end{equation}
where the $g^{(i)}=(g^{(i)}_1,\,g^{(i)}_2,\,g^{(i)}_3)\in \IZ_n=\{1, \theta, \theta^2,...\, , \theta^{n-1}\}$ such that 
\begin{equation}\label{eq:criterion}{\sum_{i=1}^3 g_i=1,\quad  0\leq g_i<1.}\end{equation}
$\theta$ always fulfills this criterion.  We denote the the exceptional divisors corresponding to the $w_i$ by $E_i$. To each of the new generators we associate a new coordinate which we denote by $y_i$, as opposed to the $z_i$ we associated to the original $v_i$.

Let us pause for a moment to think about what this method of resolution means. The obvious reason for enforcing the criterion (\ref{eq:criterion}) is that group elements which do not respect it fail to fulfill the Calabi--Yau condition: Their third component is no longer equal to one. But what is the interpretation of these group elements that do not contribute? Another way to phrase the question is: Why do not all twisted sectors contribute exceptional divisors? A closer look at the group elements shows that all those elements of the form $\frac{1}{n}\,(1,n_1,n_2)$ which fulfill (\ref{eq:criterion}) give rise to inner points of the toric diagram. Those of the form $\frac{1}{n}\,(1,0,n-1)$ lead to points on the edge of the diagram. They always fulfill (\ref{eq:criterion}) and each element which belongs to such a sub--group contributes a divisor to the respective edge, therefore there will be $n-1$ points on it. The elements which do not fulfill (\ref{eq:criterion}) are in fact anti--twists, \ie\ they have the form $\frac{1}{n}\,(n-1, n-n_1, n-n_2)$. Since the anti--twist does not carry any information which was not contained already in the twist, there is no need to take it into account separately, so also from this point of view it makes sense that it does not contribute an exceptional divisor to the resolution.

The case $\IC^2/\IZ_n$ is even simpler. The singularity $\IC^2/\IZ_n$ is called a \emph{rational double point} of type $A_{n-1}$ and its resolution is called a \emph{Hirzebruch--Jung sphere tree} consisting of $n-1$ exceptional divisors intersecting themselves according to the Dynkin diagram of $A_{n-1}$. The corresponding polyhedron $\Delta^{(1)}$ consists of a single edge joining two vertices $v_1$ and $v_2$ with $n-1$ equally spaced lattice points $w_1,\dots,w_{n-1}$ in the interior of the edge. 

Now we subdivide the cone. The diagram of the resolution of ${\IC}^3/G$ contains $n$ triangles, where $n$ is the order of $G$, yielding $n$ three--dimensional cones. For most groups $G$, several triangulations, and therefore several resolutions are possible (for large group orders even several thousands). They are all related via birational\footnote{A \emph{birational} map between algebraic varieties is a rational map with a rational inverse. A \emph{rational} map from a complex manifold $M$ to projective space $\IP^n$ is a map $f: z\to[1,f_1(z),...,f_n(z)]$ given by $n$ global meromorphic functions on $M$.} transformations, namely flop transitions. Some physical properties change for different triangulations, such as the intersection ring. Different triangulations correspond to different phases in the K\"ahler moduli space.

This treatment is easily extended to
${\IC}^3/\IZ_N\times\IZ_M$--orbifolds. When constructing the fan, the
coordinates of the generators $v_i$ not only have to fulfill one
equation (\ref{eqfan}) but three, coming from the twist $\theta^1$
associated to $\IZ_N$, the twist $\theta^2$ associated to $\IZ_M$ and
from the combined twist $\theta^1\theta^2$. When blowing up the
orbifold, the possible group elements $g^{(i)}$ are
\begin{equation}
  \{(\theta^1)^i(\theta^2)^j,\ i=0,...,N-1,\ j=0,...,M-1\}. 
\end{equation}
The toric diagram of the blown--up geometry contains $N\cdot M$
triangles corresponding to the tree--dimensional cones. The remainder
of the preceding discussion remains the same.

We also want to settle the question to which toric variety the blown--up geometry corresponds. 
Applied to our case $X_\Sigma={\IC}^3/G$, the new blown up variety corresponds to
\begin{equation}\label{eq:blowup}{X_{\tilde\Sigma}=\left({\IC}^{3+d}\setminus F_{\tilde\Sigma}\right)/({\IC}^*)^d,}\end{equation}
where $d$ is the number of new generators $w_i$ of one--dimensional cones. The action of $(\IC^*)^d$ corresponds to the set of rescalings that leave the
\begin{equation}\label{newu}{\tilde U_i=z_1^{(v_1)_i}z_2^{(v_2)_i}z_3^{(v_3)_i}(y_1)^{(w_1)_i}\!...\,(y_{d})^{(w_d)_i}}\end{equation}
invariant.
The excluded set $F_{\tilde\Sigma}$ is determined as follows: Take the set of all combinations of generators $v_i$ of one--dimensional cones in $\Sigma$ that {\it do not span a cone} in $\Sigma$ and define for each such combination a linear space by setting the coordinates associated to the $v_i$ to zero. $F_\Sigma$ is the union of these linear spaces, \ie\ the set of simultaneous zeros of coordinates not belonging to the same cone.  
In the case of several possible triangulations, it is the excluded set that distinguishes the different resulting geometries.



\begin{exa1}
\label{sec:rzsixi}
  We will now resolve the singularity of $\IC^3/\IZ_{6-I}$.  The group
  elements are $\theta=\tfrac{1}{6}(1,1,4)$,
  $\theta^2=\tfrac{1}{3}(1,1,1)$, $\theta^3=\tfrac{1}{2}(1,1,0)$,
  $\theta^4=\tfrac{1}{3}(2,2,2)$ and
  $\theta^5=\tfrac{1}{6}(5,5,2)$. $\theta,\,\theta^2$ and $\theta^3$
  fulfill condition (\ref{eq:criterion}). This leads to the following
  new generators:
  \begin{eqnarray}
    w_1&=&\tfrac{1}{6}\,v_1+\tfrac{1}{6}\,v_2+\tfrac{4}{6}\,v_3=(0,0,1),\nonumber \\[2 pt]
    w_2&=&\tfrac{1}{3}\,v_1+\tfrac{1}{3}\,v_2+\tfrac{1}{3}\,v_3=(0,-1,1),\nonumber \\[2 pt]
    w_3&=&\tfrac{1}{2}\,v_1+\tfrac{1}{2}\,v_2=(0,-2,1).
  \end{eqnarray}
  In this case, the triangulation is unique. Figure \ref{fig:fsixi}
  shows the corresponding toric diagram and its dual graph.
  \begin{figure}[h!]
    \begin{center}
      \includegraphics[width=0.75\textwidth]{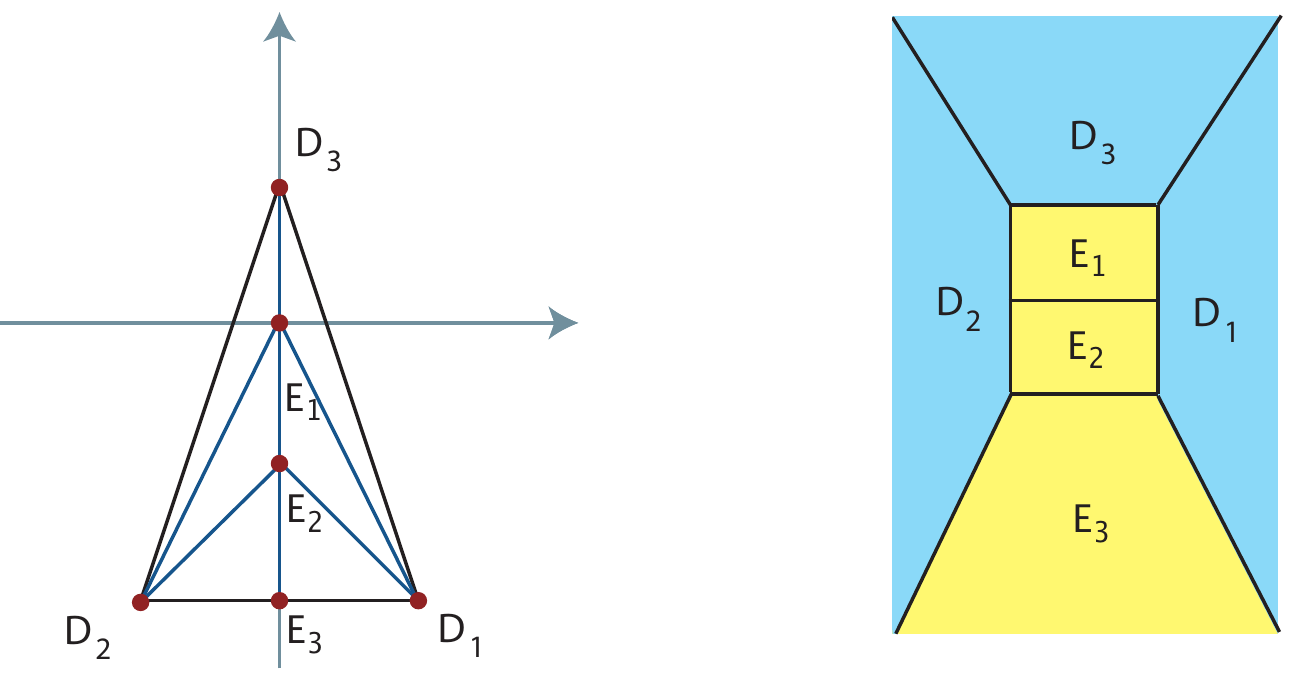}
      \caption{Toric diagram of the resolution of $\IC^3/\IZ_{6-I}$
        and dual graph}
      \label{fig:fsixi}
    \end{center}
  \end{figure}
  Let us identify the new geometry. The $\tilde U_i$ are
  \begin{eqnarray}
    \label{eq:tildeUsixi}
    \tilde U_1&=&{z_1\over z_2},\quad \tilde U_2={z_3\over z_1^2z_2^2y_2y_3^2},\quad \tilde U_3=z_1z_2z_3y_1y_2y_3.
  \end{eqnarray}
  The rescalings that leave the $\tilde U_i$ invariant are
  \begin{equation}
    \label{eq:rescalessixi}
    (z_1,\,z_2,\,z_3,\,y_1,\,y_2,\,y_3) \to (\lambda_1\,z_1,\,\lambda_1\,z_2,\,\lambda_1^4\lambda_2\lambda_3\,z_3,\, {1\over\lambda_1^6\lambda_2^2\lambda_3^3}\,y_1,\,\lambda_2\,y_2,\,\lambda_3\,y_3).
  \end{equation}
  According to eq.~(\ref{eq:blowup}), the new blown--up geometry is
  \begin{equation}
    \label{eq:blowupsixi}
    X_{\tilde\Sigma}=\,({\IC}^{6}\setminus F_{\tilde\Sigma})/({\IC}^*)^3,
  \end{equation}
  where the action of $(\IC^*)^3$ is given by
  eq.~(\ref{eq:rescalessixi}).  The excluded set is generated by
  \begin{equation*}
    \label{eq:excludesixi}
    F_{\tilde\Sigma}=\{(z_3,y_2)=0,\,(z_3,y_3)=0,\,(y_1,y_3)=0,\,(z_1,z_2)=0\,\}.
  \end{equation*}

  As can readily be seen in the dual graph, we have seven compact
  curves in $X_{\tilde\Sigma}$. Two of them, $\{y_1=y_2=0\}$ and
  $\{y_2=y_3=0\}$ are exceptional. They both have the topology of
  ${\IP}^1$. Take for example $C_1$: To avoid being on the excluded
  set, we must have $y_3\neq0,\ z_3\neq0$ and
  $(z_1,z_2)\neq0$. Therefore $C_1=\{(z_1,z_2, 1,0,0,1),
  (z_1,z_1)\neq0\}/(z_1,z_2)$, which corresponds to a ${\IP}^1$.

  We have now six three--dimensional cones: $S_1=(D_1,\,E_2,\,E_3),\
  S_2=( D_1,\,E_2,\,E_1),\ S_3=( D_1,\,E_1,\,D_3 ),$ $S_4=(
  D_2,\,E_2,\,E_3),\ S_5=( D_2,\,E_2,\,E_1),$ and $S_6=(
  D_2,\,E_1,\,D_3)$.
\end{exa1}


\begin{exb1}
\label{sec:divsixii}
  We briefly give another example to illustrate the relation between
  different triangulations of a toric diagram. The resolution of
  $\IC^3/\IZ_{6-II}$ allows five different triangulations. Figure
  \ref{fig:sixiifive} gives the five toric diagrams.

  \begin{figure}[h!]
    \begin{center}
      \includegraphics[width=0.85\textwidth]{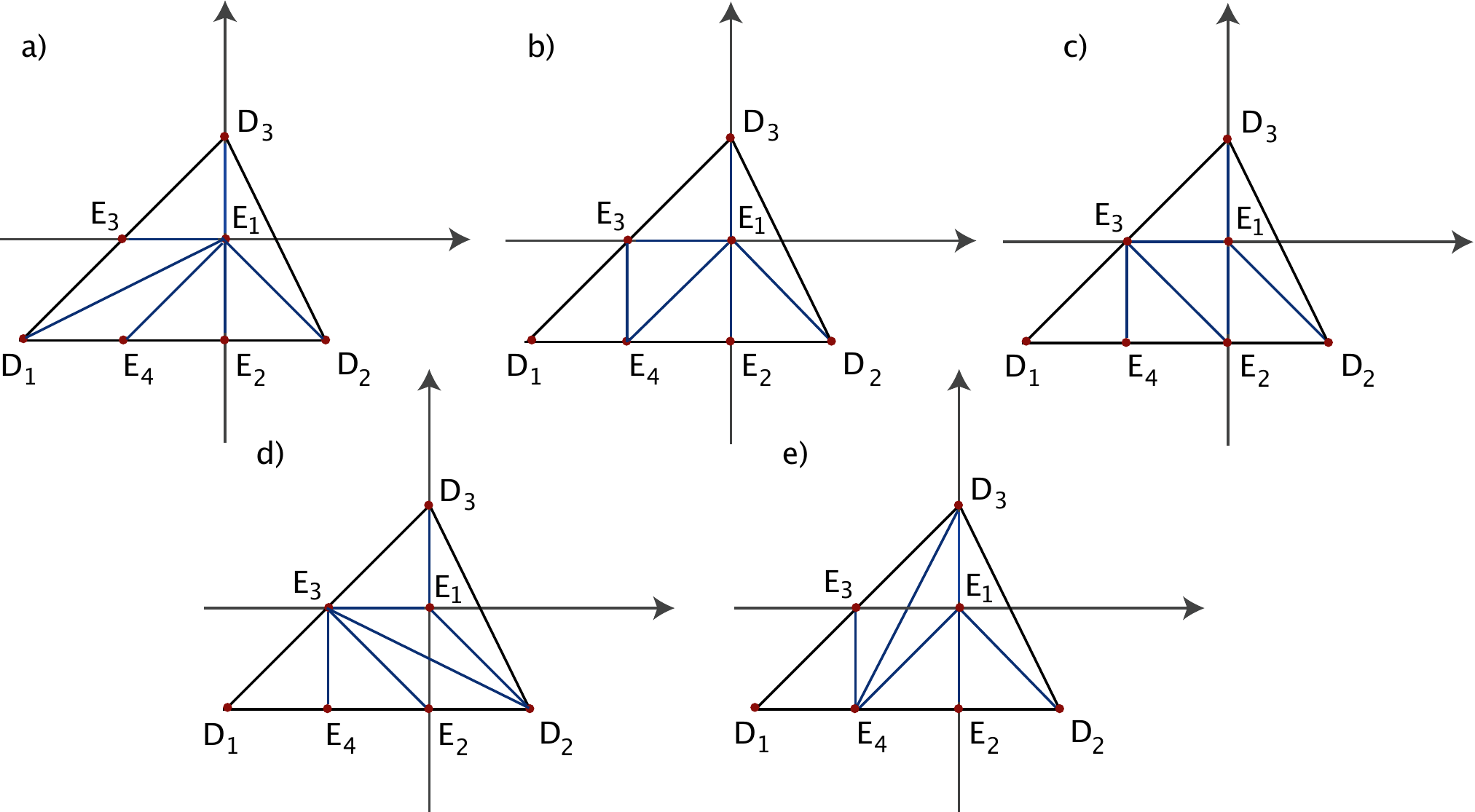}
      \caption{The five different triangulations of the toric diagram
        of the resolution of $\IC^3/\IZ_{6-II}$}
      \label{fig:sixiifive}
    \end{center}
  \end{figure}

  We start out with triangulation a). When the curve $D_1\cdot E_1$ is
  blown down and the curve $E_3\cdot E_4$ is blown up instead, we have
  gone through a flop transition and arrive at the triangulation
  b). From b) to c) we arrive by performing the flop $E_1\cdot E_4 \to
  E_2\cdot E_3$. From c) to d) takes us the flop $E_1\cdot E_2 \to
  D_2\cdot E_3$.  The last triangulation e) is produced from b) by
  flopping $E_1\cdot E_3 \to D_3\cdot E_4$. Thus, all triangulations
  are related to each other by a series of birational transformations.

\end{exb1}


\section{Mori cone and intersection numbers}\label{sec:mori}

The intersection ring of a variety is an important quantity which often proves to be of interest for the physicist (\eg\ to determine the K\"ahler potential for the K\"ahler moduli space).
The framework of toric geometry allows us to extract the desired information with ease.

Note that the intersection number of two cycles $A,\ B$ only depends on the homology classes of $A$ and $B$.
Note also that $\sum b_i D_i$ and $\sum b_i' D_i$ (where the $D_i$ are the divisors corresponding to the one--dimensional cones) are linearly equivalent if and only if they are homologically equivalent.

To arrive at the equivalences in homology, we first identify the linear relations between the divisors of the form 
\begin{equation}
  a^i_1\,v_1+a^i_2\,v_2+a^i_3\,v_3+a^i_4\,w_1+...+a^i_{3+d}\,w_d=0 \, .
\end{equation}
These linear relations can be obtained either by direct examination of the generators or can be read off directly from the algebraic torus action $({\IC}^*)^m$. The exponents of the different scaling parameters yield the coefficients $a_i$. The divisors corresponding to such a linear combination are sliding divisors in the compact geometry.  It is very convenient to introduce a matrix $(\,P\, |\,Q\,)$: The rows of $P$ contain the coordinates of the vectors $v_i$ and $w_i$. The columns of $Q$ contain the linear relations between the divisors, \ie\  the vectors $\{a^i\}$. From the rows of $Q$, which we denote by $C_i,\ i=1,...,d$, we can read off the linear equivalences in homology between the divisors which enable us to compute all triple intersection numbers. For most applications, it is most convenient to choose the $C_i$ to be the generators of the \emph{Mori cone}. The Mori cone is the space of \emph{effective curves}, \ie\ the space of all curves 
\begin{equation}
C\in X_\Sigma\ \text{with}\ C\cdot D\geq 0 \text{ for all divisors } D\in X_\Sigma.
\end{equation} 
It is dual to the K\"ahler cone. In our cases, the Mori cone is spanned by curves corresponding to two--dimensional cones. The curves correspond to the linear relations for the vertices. The generators for the Mori cone correspond to those linear relations in terms of which all others can be expressed as positive, integer linear combinations.

We will briefly survey the method of finding the generators of the Mori cone, which can be found \eg\ in \cite{Berglund:1995gd}.
\begin{itemize}
\item[I.] In a given triangulation, take the three--dimensional simplices $S_k$ (corresponding to the three--dimensional cones). Take those pairs of simplices $(S_l,S_k)$ that share a two--dimensional simplex $S_k\cap S_l$.
\item[II.] For each such pair find the unique linear relation among the vertices in $S_k\cup S_l$ such that 
\begin{itemize}
\item[(i)] the coefficients are minimal integers and 
\item[(ii)] the coefficients for the points in $(S_k \cup S_l) \setminus (S_k \cap S_l)$ are positive.
\end{itemize}
\item[III.] Find the minimal integer relations among those obtained in step 2 such that each of them can be expressed as a positive integer linear combination of them. 
\end{itemize}
While the first two steps are very simple, step III. becomes increasingly tricky for larger groups.

The general rule for triple intersections is that the intersection number of three distinct divisors is 1 if they belong to the same cone and 0 otherwise. The set of collections of divisors which do not intersect because they do not lie in the same come forms a further characteristic quantity of a toric variety, the Stanley--Reisner ideal. It contains the same information as the exceptional set $F_{\Sigma}$. Intersection numbers for triple intersections of the form $D_i^2 D_j$ or $E_k^3$ can be obtained by making use of the linear equivalences between the divisors. Since we are working here with non--compact varieties at least one compact divisor has to be involved. For intersections in compact varieties there is no such condition. The intersection ring of a toric variety is -- up to a global normalization -- completely determined by the linear relations and the Stanley--Reisner ideal. The normalization is fixed by one intersection number of three distinct divisors.

The matrix elements of $Q$ are the intersection numbers between the curves $C_i$ and the divisors $D_i,\,E_i$. We can use this to determine how the compact curves of our blown--up geometry are related to the $C_i$. 



\begin{exa1}
  \label{sec:exsixiaa}

  For this example, the method of working out the Mori generators is
  shown step by step.  We give the pairs, the sets $S_l\cup S_k$ (the
  points underlined are those who have to have positive coefficients)
  and the linear relations:
  \begin{eqnarray}
    \label{eq:Moripairs}
    1.\ \  S_6\cup S_3&=&\{\underline{D_1},\,\underline{D_2},\,D_3,\,E_1\},\quad D_1+D_2+4\, D_3-6\,E_1=0,\nonumber\\
    2.\ \   S_5\cup S_2&=&\{\underline{D_1},\,\underline{D_2},\,E_1,\,E_2\},\quad D_1+D_2+2\, E_1-4\,E_2=0,\nonumber\\
    3.\ \   S_4\cup S_1&=&\{\underline{D_1},\,\underline{D_2},\,E_2,\,E_3\},\quad D_1+D_2-2\,E_3=0,\nonumber\\
    4.\ \   S_3\cup S_2&=&\{D_1,\,\underline{D_3},\,E_1,\,\underline{E_2}\},\quad D_3-2\,E_1+E_2=0,\nonumber\\
    5.\ \   S_2\cup S_1&=&\{D_1,\,\underline{E_1},\,E_2,\,\underline{E_3}\},\quad E_1-2\,E_2+E_3=0,\nonumber\\
    6.\ \  S_6\cup S_5&=&\{D_2,\,\underline{D_3},\,E_1,\,\underline{E_2}\},\quad D_3-2\,E_1+E_2=0,\nonumber\\
    7.\ \   S_5\cup S_4&=&\{D_2,\,\underline{E_1},\,E_2,\,\underline{E_3}\},\quad E_1-2\,E_2+E_3=0.
  \end{eqnarray}
  With the relations 3, 4 and 5 all other relations can be expressed
  as a positive integer linear combination.  This leads to the
  following three Mori generators:
  \begin{equation}
    \label{eq:Morigensixi}
    C_1=\{0,0,0,1,-2,1\},\quad C_2=\{1,1,0,0,0,-2\},\quad C_3=\{0,0,1,-2,1,0\}.
  \end{equation}
  With this, we are ready to write down $(P\,|\,Q)$:
  \begin{equation}
    \label{eq:PQ}
    (P\,|\,Q) = 
      \begin{pmatrix} 
        D_1&1&-2&1&|&0&1&0 \\ 
        D_2&-1&-2&1&|&0&1&0 \\
        D_3&0&1&1&|&0&0&1 \\
        E_1&0&0&1&|&1&0&\!\!-2 \\
        E_2&0&-1&1&|&\!\!\!-2&0&1 \\
        E_3&0&-2&1&|&1&\!\!\!-2&0
      \end{pmatrix} \, .
  \end{equation}
  From the rows of $Q$, we can read off directly the linear
  equivalences:
  \begin{equation}
    D_1 \sim D_2,\quad E_2 \sim -2\,E_1 - 3\,D_3,\quad E_3 \sim E_1-2\,D_1 + 2\,D_3 \,. 
  \end{equation}
  The matrix elements of $Q$ contain the intersection numbers of the
  $C_i$ with the $D_1,\,E_1$, \eg\ $E_1\cdot C_3=-2,\ D_3\cdot C_1=0$,
  etc.  We know that $E_1\cdot E_3=0$. From the linear equivalences
  between the divisors, we find the following relations between the
  curves $C_i$ and the seven compact curves of our geometry:
  \begin{subequations}
    \begin{align}
      C_1 &= D_1\cdot E_2=D_2\cdot E_2,\\
      C_2 &= E_2\cdot E_3,\\
      C_3 &= D_1\cdot E_1=D_2\cdot E_1,\\
      E_1\cdot E_2&= 2\,C_1+C_2,\\
      D_3\cdot E_1&= 2\,C_1+C_2+4\,C_3.
    \end{align}
  \end{subequations}
  {From} these relations and $(P\,|\,Q)$, we can get all triple
  intersection numbers, \eg\
  \begin{equation}
    E_1^2E_2=E_1E_2E_3+2\,D_1E_1E_2-2\,D_3E_1E_2=2 \, .
  \end{equation}
  Table \ref{table:inter} gives the intersections of all compact curves
  with the divisors.
\begin{table}
  \begin{center}
    \begin{tabular} {ccccccc}\toprule {
        Curve}&$D_1$&$D_2$&$D_3$&$E_1$&$E_2$&$E_3$\cr
      \midrule
      \rowcolor[gray]{.95}$E_1\cdot E_2$&1&1&0 &2&\!\!\!-4&0\cr
      $E_2\cdot E_3$&1&1&0&0&0&\!\!\!-2\cr
      \rowcolor[gray]{.95}$D_1\cdot E_1$&0&0&1&\!\!\!-2&1&0\cr
      $D_1\cdot E_2$&0&0&0&1&\!\!\!-2&1\cr
      \rowcolor[gray]{.95}$D_2\cdot E_1$&0&0&1&\!\!\!-2&1&0\cr
      $D_2\cdot E_2$&0&0&0&1&\!\!\!-2&1\cr
      \rowcolor[gray]{.95}$D_3\cdot E_1$&1&1&4&\!\!\!-6&0&0\cr
      \bottomrule
    \end{tabular}
  \end{center}
  \caption{Triple intersection numbers of the blow--up of $\IZ_{6-II}$}\label{table:inter}
  \label{tab:inter}
\end{table}

Using the linear equivalences, we can also find the triple
self--intersections of the compact exceptional divisors:
\begin{equation}
  E_1^3=E_2^3=8 \, .\label{eq:3}
\end{equation}
From the intersection numbers in $Q$, we find that $\{E_1+2\,D_3, D_2,
D_3\}$ form a basis of the K\"ahler cone which is dual to the basis
$\{C_1,C_2,C_3\}$ of the Mori cone.

\end{exa1}


\section{Divisor topologies}\label{sec:ExcepTop}

There are two types of exceptional divisors: The compact divisors, whose corresponding points lie in the interior of the toric diagram, and the semi--compact ones whose points sit on the boundary of the toric diagram. The latter case corresponds to the two--dimensional situation with an extra non--compact direction, hence it has the topology of $\IC \times \IP^1$ with possibly some blow--ups.
The $D$--divisors are non--compact and of the form $\IC^2$.

We first discuss the compact divisors. For this purpose we use the notion of the \emph{star} of a cone $\sigma$, in terms of which the topology of the corresponding divisor is determined. The star, denoted $\mathrm{ Star}(\sigma)$ is the set of all cones $\tau$ in the fan $\Sigma$ containing $\sigma$. This means that we simply remove from the fan $\Sigma$ all cones, \ie\ points and lines in the toric diagram, which do not contain $w_i$. The diagram of the star is not necessarily convex anymore. Then we compute the linear relations and the Mori cone for the star. This means in particular that we drop all the simplices $S_k$ in the induced triangulation of the star which do not lie in its toric diagram. As a consequence, certain linear relations of the full diagram will be removed in the process of determining the Mori cone.  The generators of the Mori cone of the star will in general be different from those of $\Sigma$. Once we have obtained the Mori cone of the star, we can rely on the classification of compact toric surfaces. 

\begin{figure}
  \centering
  \hfill \subfigure[fan of the Hirzebruch surface $\IF_n$]{\includegraphics[width=.3\textwidth]{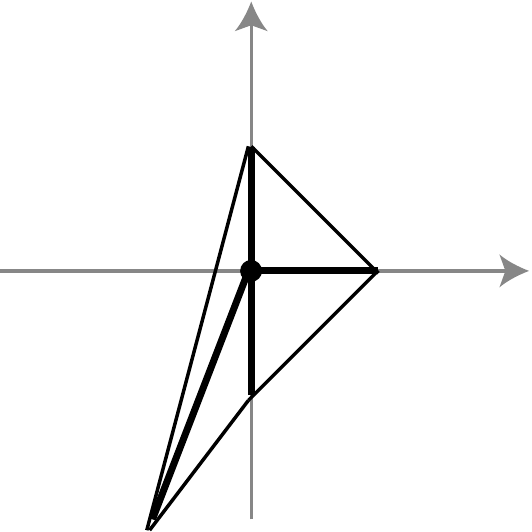}} 
  \hfill \subfigure[fan of $dP_0=\IP^2$]{\includegraphics[width=.3\textwidth]{P2}} \hfill
  \subfigure[fan of $\IP^1\times\IP^1=\IF_0$]{\includegraphics[width=.3\textwidth]{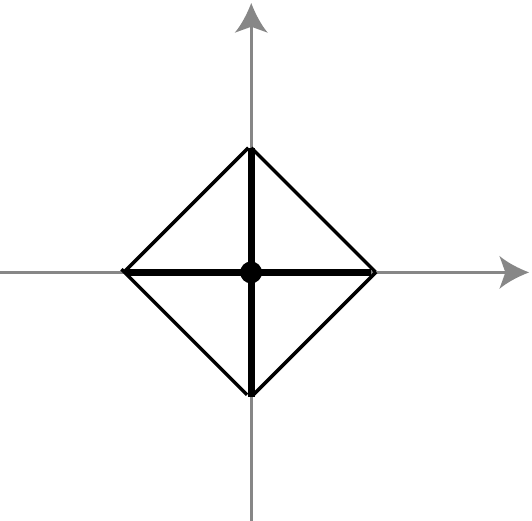}} \vfill  
  \hfill\subfigure[fan of $dP_1=\mathrm{Bl}_1\IP^2$]{\includegraphics[width=.3\textwidth]{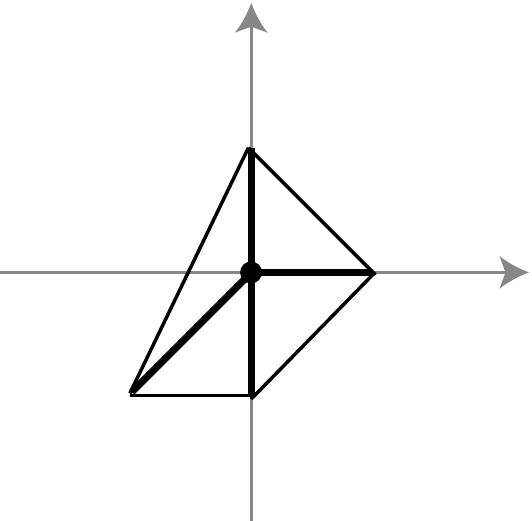}}  
  \hfill\subfigure[fan of $dP_2=\mathrm{Bl}_2\IP^2$]{\includegraphics[width=.3\textwidth]{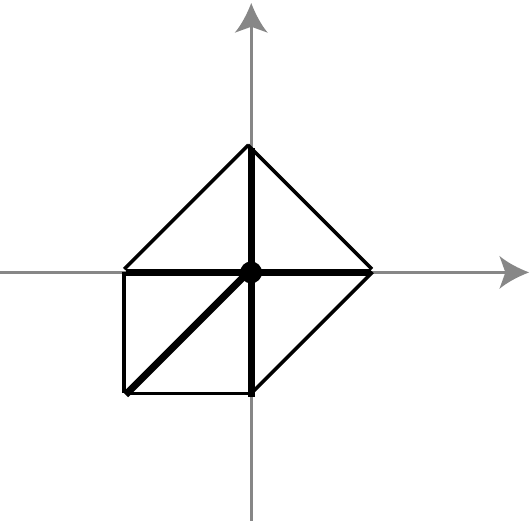}}  
  \hfill  \subfigure[fan of $dP_3=\mathrm{Bl}_3\IP^2$]{\includegraphics[width=.3\textwidth]{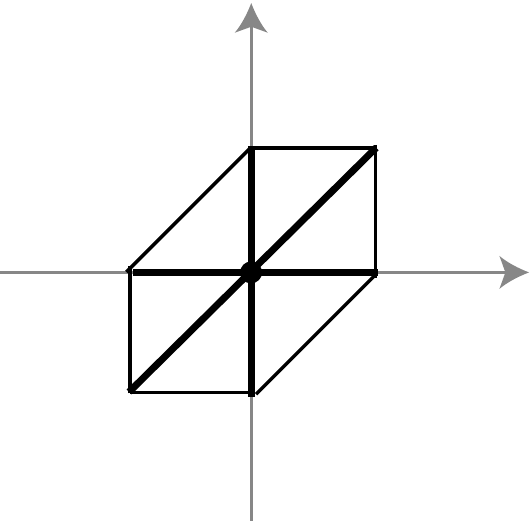}} \hfill
  \caption{Fans of $\IF_n$ and the toric del Pezzo surfaces}
  \label{fig:delPezzo}
\end{figure}

\begin{framed}
  \begin{minipage}{.85\textwidth}
    \subsection*{Digression: Classification of compact toric surfaces}

    Any toric surface is either a $\IP^2$, a Hirzebruch surface
    $\IF_n$, or a toric blow--up thereof. The simplest possible
    surface is obviously $\IP^2$. Each surface, which is birationally
    equivalent to $\IP^2$ is called a \emph{rational surface}.

    \subsubsection{Hirzebruch surfaces}

    A \emph{Hirzebruch}\footnote{Friedrich E.P. Hirzebruch (*1927),
      German mathematician} \emph{surface} $\mathbb{F}_n$ is a special
    case of a \emph{ruled surface} $S$, which admits a fibration
    \begin{equation*}
      \pi:\,S\to C,
    \end{equation*}
    $C$ a smooth curve and the generic fiber of $\pi$ being isomorphic
    to $\IP^1$.  A Hirzebruch surface is a fibration of $\mathbb{P}^1$
    over $\mathbb{P}^1$ and is of the form
    $\IF_n=\IP(\bun_{\IP^1}\oplus\bun_{\IP^1}(n))$.

    \subsubsection{del Pezzo surfaces}

    A \emph{del Pezzo}\footnote{Pasquale del Pezzo (1859-1936),
      Neapolitan mathematician} \emph{surface} is a two--dimensional
    \emph{Fano variety}, \ie\ a variety whose anticanonical bundle is
    ample\footnote{A line bundle $L$ is \emph{very ample}, if it has
      enough sections to embed its base manifold into projective
      space. $L$ is \emph{ample}, if a tensor power $L^{\otimes n}$ of
      $L$ is very ample.}.

    In total, there exist 10 of them: $dP_0=\IP^2$,
    $\IP^1\times\IP^1=\IF_0$ and blow--ups of $\IP^2$ in up to 8
    points,
    \begin{equation*}
      \mathrm{Bl}_n\IP^2=dP_n.
    \end{equation*}
    Five of them are realized as toric surfaces, namely $\IF_0$ and
    $dP_n,\, n=0,...,3$. The fans are given in
    Figure~\ref{fig:delPezzo}. In Figure~\ref{fig:delPezzo}.a, the fan
    of $\IF_n$ is shown.
\end{minipage}
\end{framed}
\bigskip
The generator of the Mori cone of $\IP^2$ has the form 
\begin{equation}
  Q^T = \left(\begin{array}{cccc}-3& 1& 1& 1\end{array}\right) \, .
\end{equation}
For $\IF_n$, the generators take the form
\begin{equation}
  Q^T =
  \left(\begin{array}{ccccc}-2 & 1 & 1 & 0 & 0\cr-n-2 & 0 & n & 1 &
      1\end{array}\right) \qquad \text{ or} \qquad Q^T =
  \left(\begin{array}{ccccc} -2 & 1 & 1 & 0 & 0\cr n-2 & 0 & -n & 1 &
      1\end{array}\right)
\end{equation}
since $\IF_{-n}$ is isomorphic to $\IF_n$. Finally, every toric
blow--up of a point adds an additional independent relation whose form
is 
\begin{equation}
  Q^T = \left(\begin{array}{cccccc}0 & ... & 0 &1 &1
      &-2\cr\end{array}\right) \, .
\end{equation}
We will denote the blow--up of a surface $S$ in $n$ points by
$\mathrm{Bl}_{n}S$.  The toric variety $X_\Sigma$ is
three dimensional, which means in particular that the stars are in
fact cones over a polygon. An additional possibility for a toric
blow--up is adding a point to the polygon such that the corresponding
relation is of the form
\begin{equation}
  Q^T = \left(\begin{array}{ccccccc}0 & ... & 0 &1 &1 &-1 &-1\end{array}\right) \, .
\end{equation}
This corresponds to adding a cone over a lozenge and is well--known
from the resolution of the conifold singularity.

Also the semi--compact exceptional divisors can be dealt with using
the star. Since the geometry is effectively reduced by one dimension,
the only compact toric manifold in one dimension is $\IP^1$ and the
corresponding generator is
\begin{equation}
  Q^T = \left(\begin{array}{cccc}-2& 1& 1&
      0\cr\end{array}\right) \, ,
\end{equation}
where the 0 corresponds to the non--compact factor $\IC$.



\begin{exa1}
  \label{sec:exsixiaaa}

  We now determine the topology of the exceptional divisors for our
  example $\IC^3/\IZ_{6-I}$.
  \begin{figure}[h!]
    \begin{center}
      \includegraphics[width=140mm]{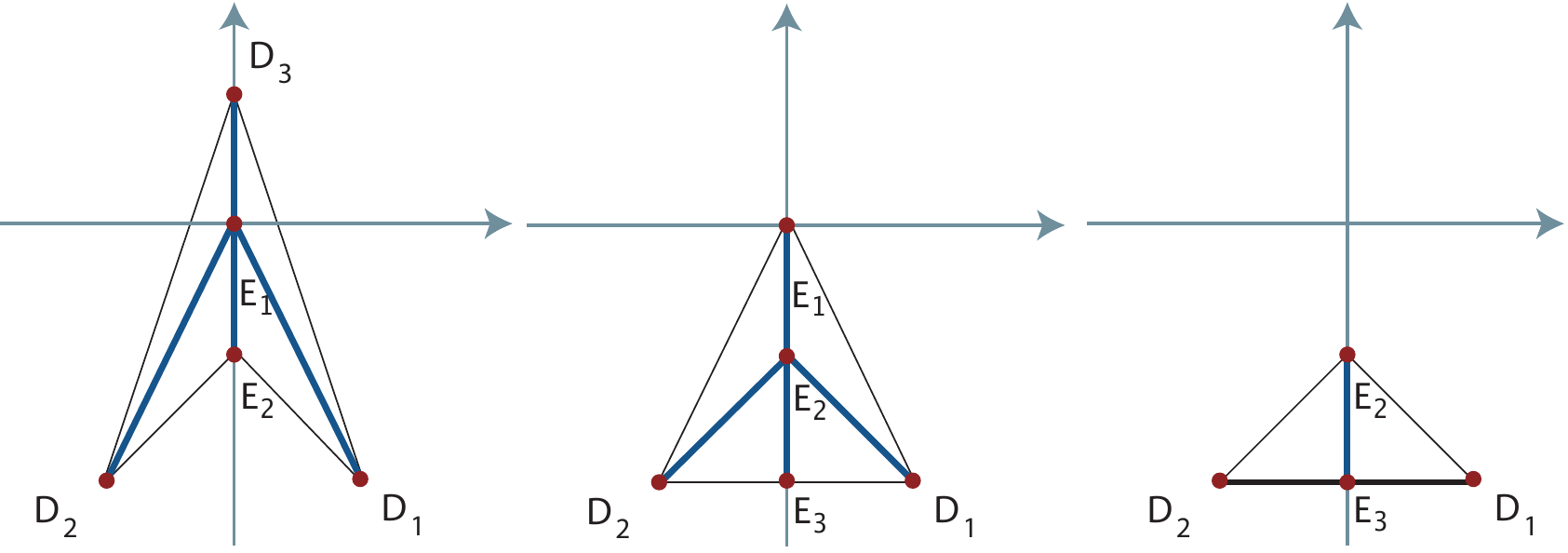}
      \caption{The stars of the exceptional divisors $E_1$, $E_2$, and
        $E_3$, respectively.}
      \label{fig:fstarsixi}
    \end{center}
  \end{figure}
  As explained above, we need to look at the respective stars which
  are displayed in Figure~\ref{fig:fstarsixi}.  In order to determine
  the Mori generators for the star of $E_1$, we have to drop the cones
  involving $E_3$ which are $S_1$ and $S_4$. From the seven relations
  in~(\ref{eq:Moripairs}) only four remain, those corresponding to
  $C_3$, $2\,C_1 + C_2$ and $2\,C_1+C_2+4\,C_3$. These are generated
  by
  \begin{eqnarray}
    2\,C_1+C_2&=&(1,1,0,2,-4,0) \text{ and }\\
    C_3&=&(0,0,1,-2,1,0),
  \end{eqnarray} 
  which are the Mori generators of $\IF_4$. Similarly, for the star of
  $E_2$ only the relations not involving $S_3$ and $S_6$ remain. These
  are generated by $C_1$ and $C_2$, and using~(\ref{eq:Morigensixi})
  we recognize them to be the Mori generators of $\IF_2$. Finally, the
  star of $E_3$ has only the relation corresponding to $C_3$. Hence,
  the topology of $E_3$ is $\IP^1\times \IC$, as it should be, since
  the point sits on the boundary of the toric diagram of $X_\Sigma$
  and no extra exceptional curves end on it.
\end{exa1}

\section{Literature}

For a first acquaintance with toric geometry, Chapter 7 of \cite{Hori} is well suited. Also \cite{Aspinwall:1993nu} contains a very readable introduction. The classical references on toric geometry are the books by Fulton \cite{Fulton} and Oda \cite{Oda}. Unfortunately, they are both not very accessible to the physicist. The reference for general techniques in algebraic geometry is \cite{Griffiths}.

A number of reviews of topological string theory briefly introduce toric geometry, such as \cite{Neitzke:2004ni, Marino:2004uf}, but from a different point of view.

\bibliographystyle{JHEP}
\bibliography{LectureReferences}


\chapter{Application: Desingularizing toroidal orbifolds}
\minitoc
\bigskip
In this lecture, I will discuss the desingularization of toroidal orbifolds employing the methods treated so far. First, I explain how to glue together the resolved toric patches to obtain a smooth Calabi--Yau manifold from the singular orbifold quotient $T^6/\Gamma$. Next, the divisors inherited directly from the covering space $T^6$ are discussed. In the following section, the full intersection ring of the smooth manifold is calculated, and lastly, the topologies of the appearing divisor classes are determined.

\section{Gluing the patches}

In the easy cases, say in the prime orbifolds $\IZ_3$ and $\IZ_7$, it is obvious how the smooth manifold is obtained: Just put one resolved patch in the location of every fixed point and you are finished. Since these patches only have internal points, the corresponding exceptional divisors are compact, hence cannot see each other, and no complications arise from gluing.

Fixed lines which do not intersect any other fixed lines and on top of which no fixed points sit also pose no problem.

But what happens, when we have fixed lines on top of which fixed points are sitting? As discussed in Section~\ref{sec:resolve}, such a fixed point already knows it sits on a fixed line, since on the edge of the toric diagram of its resolution is the number of exceptional divisors appropriate to the fixed line the point sits on top of. Internal exceptional divisors are unproblematic in this case as well, since they do not feel the global surrounding. The exceptional divisors on the edges are identified or glued together with those of the corresponding resolved fixed lines.

The larger the order of the group, the more often it happens that a point or line is fixed under several group elements. How are we to know which of the patches we should use?

In the case of fixed lines answer is: use the patch that belongs to the generator of the largest subgroup under which the patch is fixed, because the line is fixed under the whole sub-group and its exceptional divisors already count the contributions from the other group elements. For fixed points, the question is a little more tricky. One possibility is to count the number of group elements this point is fixed under, not counting anti--twists and elements that generate fixed lines. Then choose the patch with the matching number of interior points. The other possibility is to rely on the schematic picture of the fixed set configuration and choose the patch according to the fixed lines the fixed point sits on. Isolated fixed points correspond to toric diagrams with only internal, compact exceptional divisors. When the fixed point sits on a fixed line of order $k$, its toric diagram has $k-1$ exceptional divisors on one of its boundaries. If the fixed point sits at the intersection of two (three) fixed lines, it has the appropriate number of exceptional divisors on two (three) of its boundaries. The right number of interior points together with the right number of exceptional divisors sitting on the edges uniquely determines the correct patch. 

\begin{figure}[h!]
\begin{center}
\includegraphics[width=0.35\textwidth]{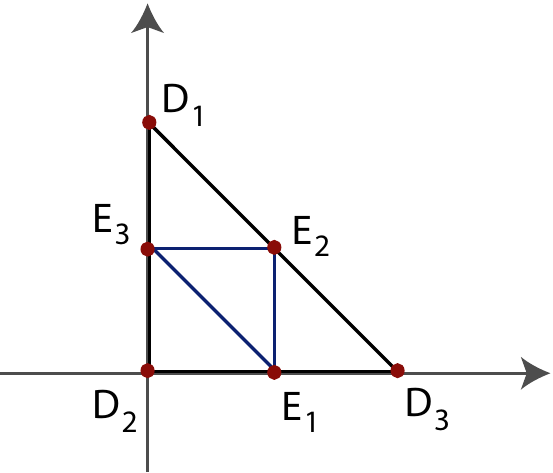}
\caption{Toric diagram of resolution of ${\IC}^3/\IZ_{2}\times\IZ_2$ and dual graph}\label{frtwotwo}
\end{center}
\end{figure}

Even though the intersection points of three $\IZ_2$ fixed lines are not fixed under a single group element, they must be resolved. The resolution of such a point is the resolution of $\IC^3/\IZ_2\times \IZ_2$ and its toric diagram is indeed the only one without interior points, see Figure~\ref{frtwotwo}.

Interestingly, the case of three intersecting $\IZ_2$ fixed lines is the only instance of intersecting fixed lines where the intersection point itself is not fixed under a single group element. This case arises only for $\IZ_n\times\IZ_m$ orbifolds with both $n$ and $m$ even.


\begin{exa}
  \label{sec:exsixiglue}

  This example is rather straightforward. We must again use the data
  of Table \ref{tab:fssixi} and the schematic picture of the fixed set
  configuration \ref{fig:ffixedi}. Furthermore, we need the resolved
  patches of $\IC^3/\IZ_{6-I}$ (see Section \ref{sec:rzsixi}, in
  particular Figure \ref{fig:fsixi}), $\IC^3/\IZ_{3}$ (see Figure
  \ref{fig:frthree}), and the resolution of the $\IZ_2$ fixed line.
  The three $\IZ_6$--patches contribute two exceptional divisors each:
  $E_{1,\gamma}$, and $E_{2,1,\gamma}$, where $\gamma=1,2,3$ labels
  the patches in the $z^3$--direction.  The exceptional divisor $E_3$
  on the edge is identified with the one of the resolved fixed line
  the patch sits upon, as we will see.
  \begin{figure}[h!]
    \begin{center}
      \includegraphics[width=.35\textwidth]{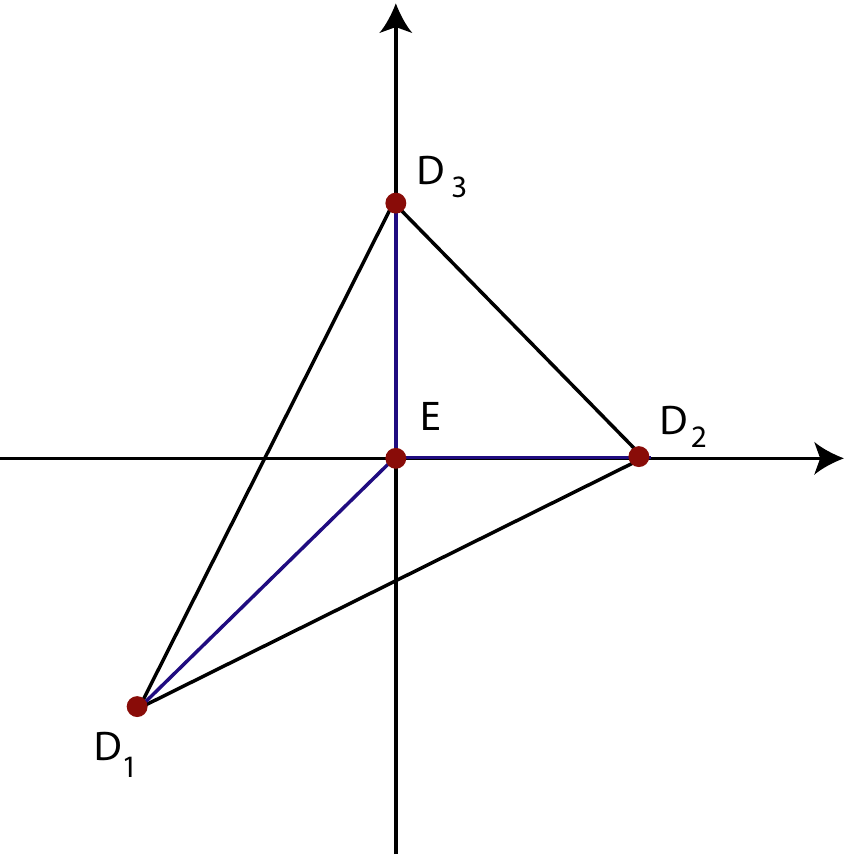}
      \caption{Toric diagram of the resolution of ${\IC}^3/\IZ_{3}$}
      \label{fig:frthree}
    \end{center}
  \end{figure}
  There are furthermore 15 conjugacy classes of $\IZ_3$ fixed
  points. Blowing them up leads to a contribution of one exceptional
  divisor as can be seen from Figure~\ref{fig:frthree}. Since three of
  these fixed points sit at the location of the $\IZ_{6-I}$ fixed
  points which we have already taken into account ($E_{2,1,\gamma}$),
  we only count 12 of them, and denote the resulting divisors by
  $E_{2,\mu,\gamma},\ \mu = 2,\dots,5, \, \gamma=1,2,3$. The invariant
  divisors are built according to the conjugacy classes, \eg
  \begin{equation}
    E_{2,2,\gamma} = \Et_{2,1,2,\gamma} +  \Et_{2,1,3,\gamma} \, ,
  \end{equation}
  etc., where $\Et_{2,\alpha,\beta,\gamma}$ are the representatives on
  the cover.  Finally, there are 6 conjugacy classes of fixed lines of
  the form $\IC^2/\IZ_2$. We see that after the resolution, each class
  contributes one exceptional divisor $E_{3,\alpha}, \alpha=1,2$. On
  the fixed line at $\zf{1}{1}=\zf{2}{1}=0$ sit the three $\IZ_{6-I}$
  fixed points. The divisor coming from the blow--up of this fixed
  line, $E_{3,1}$, is identified with the three exceptional divisors
  corresponding to the points on the boundary of the toric diagram of
  the resolution of $\IC^3/\IZ_{6-I}$ that we mentioned above.  In
  total, this adds up to
  \begin{equation}
    h^{1,1}_{twisted}=3\cdot2+12\cdot1+6\cdot1=24      
  \end{equation}
  exceptional divisors, which is the number which is given for
  $h^{(1,1)}_{twisted}$ in Table \ref{table:one}.
\end{exa}

\begin{exc}
  \label{sec:exsixsix}

  This, being the point group of largest order, is the most tedious of
  all examples. It is presented here to show that the procedure is not
  so tedious after all.

  First, the fixed sets must be identified. Table~\ref{fssixsix}
  summarizes the results.
  \begin{table}[h!]\begin{center}
      \begin{tabular}{cccc}
        \toprule
        Group el.& Order &Fixed Set& Conj. Classes \cr
        \midrule
        \rowcolor[gray]{.95}$ \theta^1$&6    &1\ { fixed\ line} &\ 1\cr
        $ (\theta^1)^2$&3        &9\ { fixed\ lines} &\ 4\cr
        \rowcolor[gray]{.95}$ (\theta^1)^3$&2       &16\ { fixed\ lines} &\ 4\cr
        $ \theta^2$&6     &1\ { fixed\ line} &\ 1\cr
        \rowcolor[gray]{.95}$ (\theta^2)^2$&3     &9 \ { fixed\ lines} &\ 4\cr
        $ (\theta^2)^3$&2    &16 \ { fixed\ lines} &\ 4\cr
        \rowcolor[gray]{.95}$ \theta^1\theta^2$&$6\times6 $   &3\ { fixed\ points} &\ 2\cr
        $ \theta^1(\theta^2)^2$&$6\times 3$    &12\ { fixed\ points} &\ 4 \cr
        \rowcolor[gray]{.95}$ \theta^1(\theta^2)^3$&$6\times2  $  &12\ { fixed\ points} &\ 4\cr
        $ \theta^1(\theta^2)^4$&$6\times6  $ &3\ { fixed\ points} &\ 2\cr
        \rowcolor[gray]{.95}$ \theta^1(\theta^2)^5$&6  &1\ { fixed\ line} &\ 1\cr
        $ (\theta^1)^2\theta^2$&$3\times6$   &12\ { fixed\ points} &\ 4\cr
        \rowcolor[gray]{.95}$(\theta^1)^3\theta^2$&$2\times6 $   &12\ { fixed\ points} &\ 4\cr
        $ (\theta^1)^4\theta^2$&$6\times6 $     &3\ { fixed\ points} &\ 2\cr
        \rowcolor[gray]{.95}$ (\theta^1)^2(\theta^2)^2$&$3\times3$  &27\ { fixed\ points} &\ 9\cr
        $ (\theta^1)^2(\theta^2)^3$&$3\times2$  &12\ { fixed\ points} &\ 4\cr
        \rowcolor[gray]{.95}$ (\theta^1)^2(\theta^2)^4$&{3} &9\ { fixed\ lines} &\ 4\cr
        $ (\theta^1)^3(\theta^2)^2$&$2\times3$  &12\ { fixed\ points} &\ 4\cr
        \rowcolor[gray]{.95}$ (\theta^1)^3(\theta^2)^3$&{2}    &16\ { fixed\ lines} &\ 4\cr
        \bottomrule
      \end{tabular}
      \caption{Fixed point set for $\IZ_6\times
        \IZ_6$.}\label{fssixsix}
    \end{center}\end{table}
  \begin{figure}[p]
    \begin{center}
      \includegraphics[height=0.9\textheight]{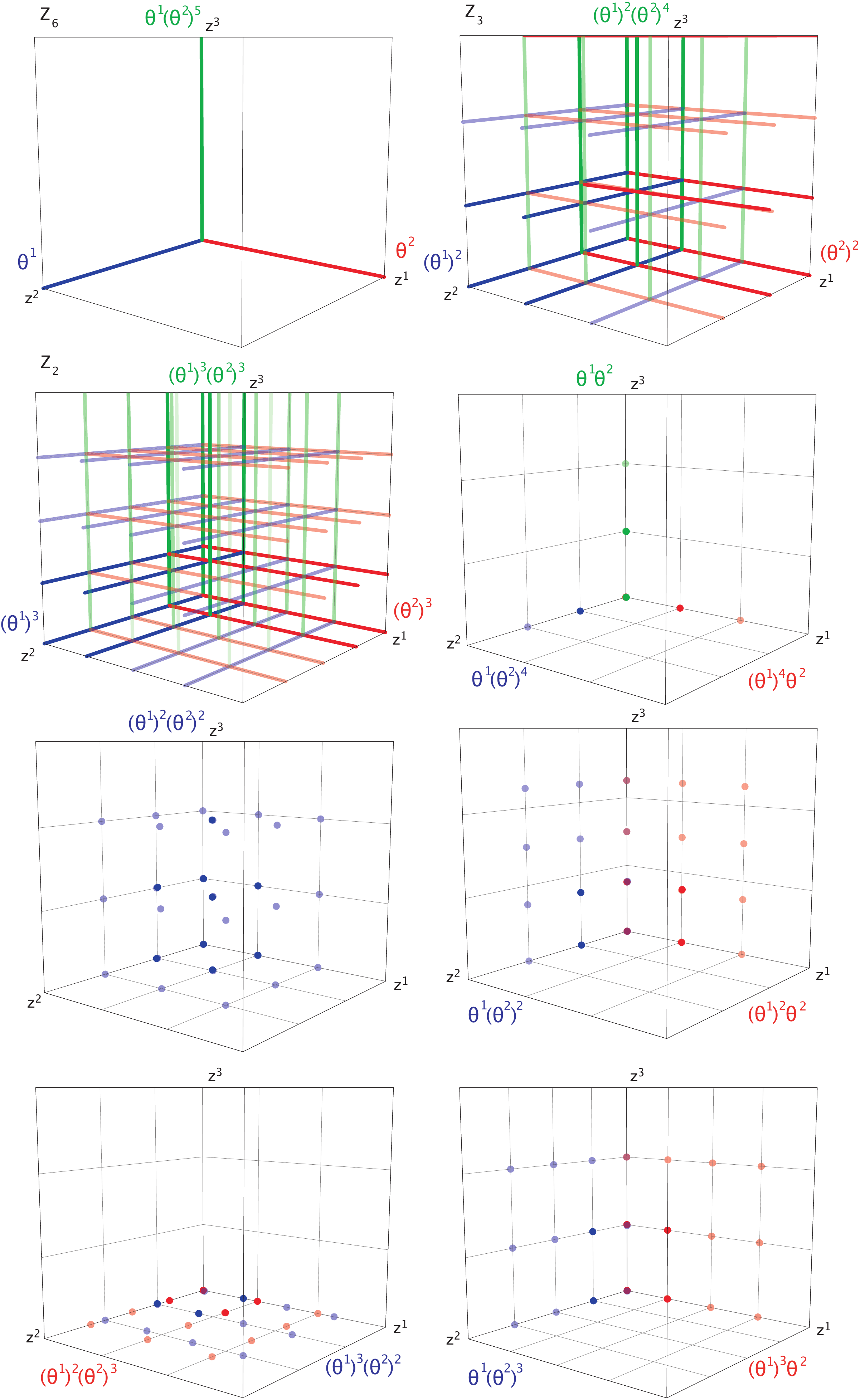}
      \caption{Schematic picture of the fixed set configuration of
        $\IZ_6\times\IZ_6$}\label{ffixsixsix}
    \end{center}
  \end{figure}
  Figure~\ref{ffixsixsix} shows the schematic picture of the fixed set
  configuration. Again, it is the covering space that is shown, the
  representatives of the equivalence classes are highlighted.

  Now we are ready to glue the patches together. Figure
  \ref{ingredients} schematically shows all the patches that will be
  needed in this example.
  \begin{figure}[h!]
    \begin{center}
      \includegraphics[width=0.55\textwidth]{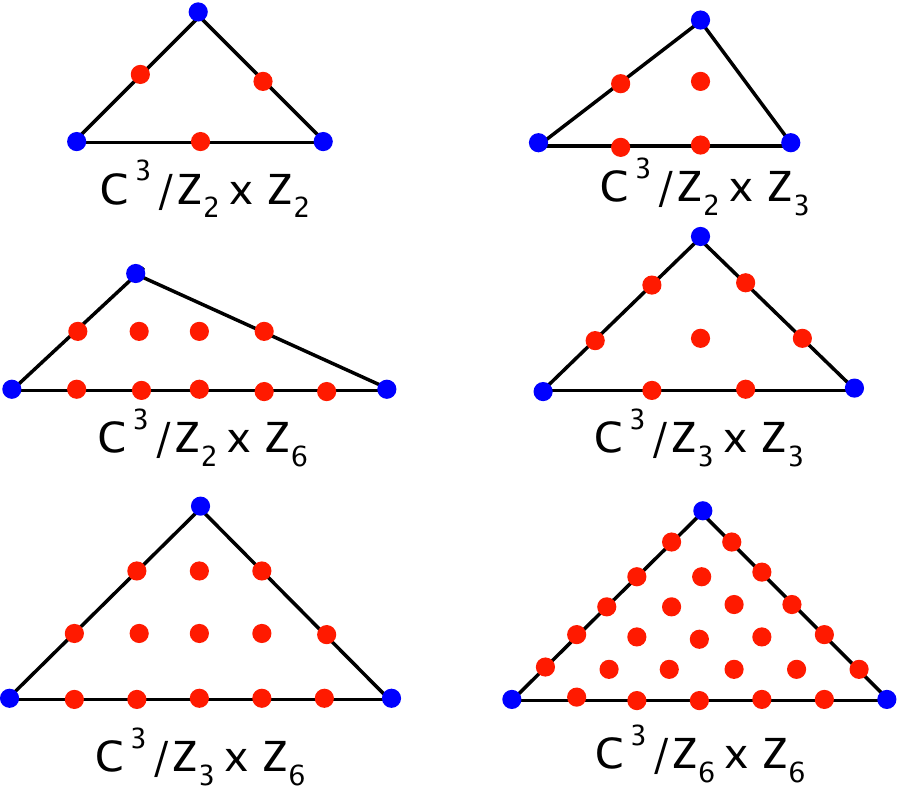}
      \caption{Toric diagrams of patches for
        $T^6/\IZ_6\times\IZ_6$}\label{ingredients}
    \end{center}
  \end{figure}
  It is easiest to first look at the fixed lines. There are three
  $\IZ_6$ fixed lines, each contributing five exceptional
  divisors. Then there are twelve equivalence classes of $\IZ_3$ fixed
  lines, three of which coincide with the $\IZ_6$ fixed lines. The
  latter need not be counted, since they are already contained in the
  divisor count of the $\IZ_6$ fixed lines. The $\IZ_3$ fixed lines
  each contribute two exceptional divisors. Furthermore, there are
  twelve equivalence classes of $\IZ_2$ fixed lines, three of which
  again coincide with the $\IZ_6$ fixed lines. They give rise to one
  exceptional divisor each. From the fixed lines originate in total
  \begin{equation}
    h^{1,1}_{lines}=3\cdot5+(12-3)\cdot2+(12-3)\cdot1=42
  \end{equation}
  exceptional divisors.

  Now we study the fixed points. We associate the patches to the fixed
  points according to the intersection of fixed lines on which they
  sit. The exceptional divisors on the boundaries of their toric
  diagrams are identified with the divisors of the respective fixed
  lines. There is but one fixed point on the intersection of three
  $\IZ_6$ fixed lines. It is replaced by the resolution of the
  $\IC^2/\IZ_6\times\IZ_6$ patch, which contributes ten compact
  internal exceptional divisors. There are three equivalence classes
  of fixed points on the intersections of one $\IZ_6$ fixed line and
  two $\IZ_3$ fixed lines. They are replaced by the resolutions of the
  $\IC^2/\IZ_3\times\IZ_6$ patch, which contribute four compact
  exceptional divisors each.  Then, there are five equivalence classes
  of fixed points on the intersections of three $\IZ_3$ fixed
  lines. They are replaced by the resolutions of the
  $\IC^2/\IZ_3\times\IZ_3$ patch, which contribute one compact
  exceptional divisor each. Furthermore, there are three equivalence
  classes of fixed points on the intersections of one $\IZ_6$ fixed
  line and two $\IZ_2$ fixed lines. They are replaced by the
  resolutions of the $\IC^2/\IZ_2\times\IZ_6$ patch, which contribute
  two compact exceptional divisors each. The rest of the fixed points
  sit on the intersections of one $\IZ_2$ and one $\IZ_3$ fixed
  line. There are six equivalence classes of them. They are replaced
  by the resolutions of the $\IC^2/\IZ_2\times\IZ_3$ patch, which is
  the same as the $\IC^2/\IZ_{6-II}$ patch, which contribute one
  compact exceptional divisor each. On the intersections of three
  $\IZ_2$ fixed lines sit resolved $\IC^2/\IZ_2\times\IZ_2$ patches,
  but since this patch has no internal points, it does not contribute
  any exceptional divisors which were not already counted by the fixed
  lines. The fixed points therefore yield
  \begin{equation}
    h^{1,1}_{pts}=1\cdot10+3\cdot4+5\cdot1+3\cdot2+6\cdot1=39
  \end{equation}
  exceptional divisors. From fixed lines and fixed points together we
  arrive at 
  \begin{equation}
  h^{1,1}_{twisted}=42+39=81
  \end{equation} exceptional divisors.
\end{exc}

\section{The inherited divisors}

So far, we have mainly spoken about the exceptional divisors which arise from the blow--ups of the singularities. In the local patches, the other natural set of divisors are the $D$--divisors, which descend from the local coordinates $\tilde z^i$ of the $\IC^3$--patch. On the compact space, \ie\ the resolution of $T^6/\Gamma$, the $D$s are not the natural quantities anymore. The natural quantities are the divisors $R_i$ which descend from the covering space $T^6$ and are dual to the untwisted $(1,1)$--forms of the orbifold.
The three forms 
\begin{equation}
  dz^i\wedge d\ov z^i,\ i=1,2,3
\end{equation}
are invariant under all twists. For each pair $n_i=n_j$ in the twist (\ref{cplxtwist}), the forms 
\begin{equation}
  dz^i\wedge d\ov z^j\,\text{ and }\,dz^j\wedge d\ov z^i
\end{equation}
are invariant as well.
 
The inherited divisors $R_i$ together with the exceptional divisors $E_{k,\alpha,\beta,\gamma}$ form a basis for the divisor classes of the resolved orbifold. 

The $D$--divisors, which in the local patches are defined by $\tilde z^i=0$ are in the compact manifold defined by
\begin{equation}\label{defD}
D_{i\alpha} = \{ z^i = \zf{i}{\alpha} \},
\end{equation}
where $\alpha$ runs over the fixed loci in the $i$th direction. Therefore, they correspond to planes localized at the fixed points in the compact geometry.

The three "diagonal" $R_i$ dual to $dz^i\wedge d\ov z^i,\ i=1,2,3$
correspond to fixed planes parallel to the $D$s which can sit
everywhere \emph{except} at the loci of the fixed points. They are
defined as
\begin{equation}
  R_i=\{ z^i = c \not = \zf{i}{\alpha} \} 
\end{equation}
and are "sliding" divisors in the sense that they can move away from the fixed point. $c$ corresponds to their position modulus.  
We need, however, to pay attention whether we use the local coordinates $\tilde z^i$ near the fixed point on the orbifold or the local coordinates $z^i$ on the cover. Locally, the map is $\tilde z^i = \left(z^i\right)^{n_i}$, where $n_i$ is the order of the group element that fixes the plane $D_i$. 
On the orbifold, the $R_i\ ,i=1,2,3$ are defined as
\begin{equation}\label{defR}
  R_i=\{ \tilde z^i = c^{n_i} \},\quad c\neq \zf{i}{\alpha}.
\end{equation}
On the cover, they lift to a union of $n_i$ divisors 
\begin{equation}
  R_i = \bigcup_{k=1}^{n_i} \{ z^i = \varepsilon^k c\} \text{ with } \varepsilon^{n_i} =1 \, .
\end{equation}

\begin{figure}[h!]
\begin{center}
\includegraphics[width=0.55\textwidth]{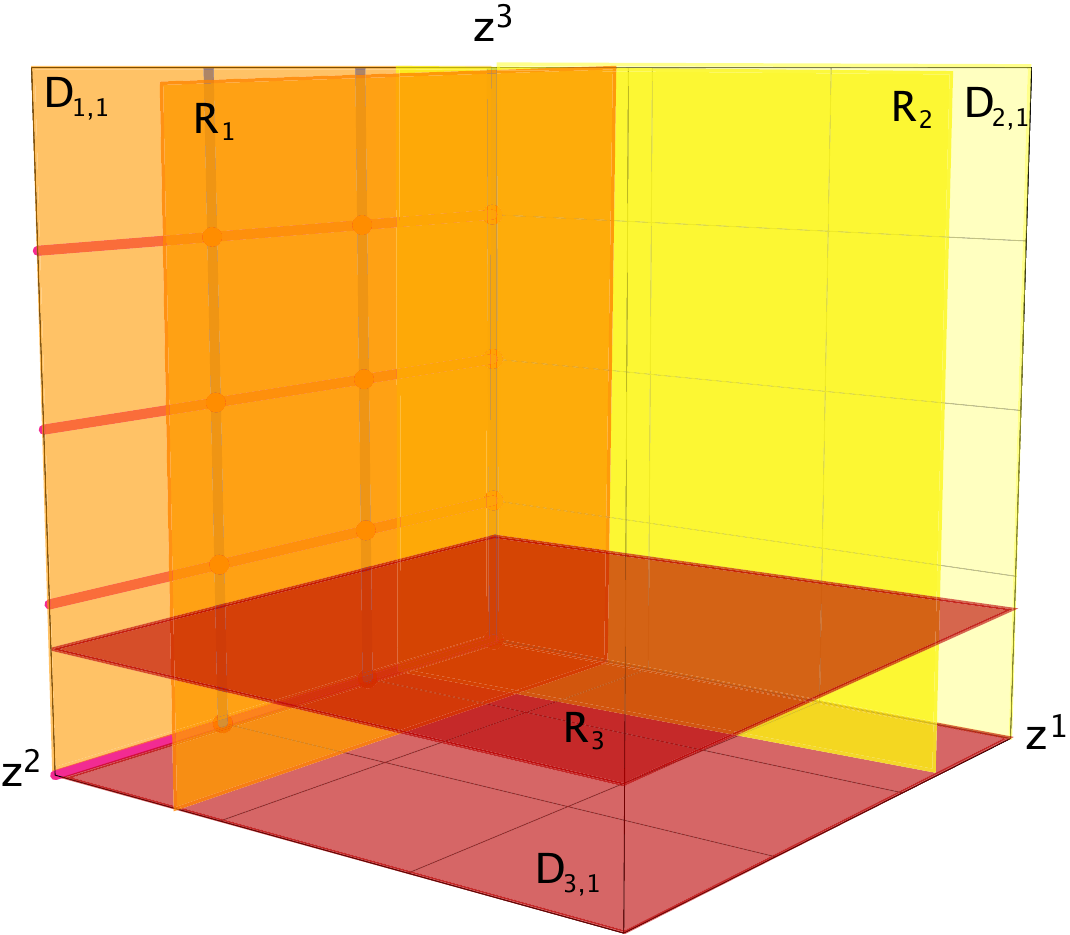}
\caption{Schematic picture of $D$- and $R$-divisors}
\label{fig:divisors}
\end{center}
\end{figure}
Figure \ref{fig:divisors} shows the schematic representation of three of the $D$ divisors and the three diagonal inherited divisors $R_i$. The figure shows the fixed set of $\IZ_{6-II}$ on $SU(2)\times SU(6)$, but this is not essential.


To relate the $R_i$ to the $D_i$, consider the local toric patch before blowing up. The fixed point lies at $c=\zf{i}{\alpha}$ and in the limit as $c$ approaches this point we find 
\begin{equation}
  R_i \sim n_i\,D_i \, . 
\end{equation}
This expresses the fact that the polynomial defining $R_i$ on the cover has a zero of order $n_i$ on $D_i$ at the fixed point. In the local toric patch $R_i \sim 0$, hence $n_i\,D_i \sim 0$. After blowing up, $R_i$ and $n_i\,D_i$ differ by the exceptional divisors $E_k$ which appear in the process of resolution. The difference is expressed precisely by the linear relation in the $i$th direction~(\ref{eq:linrels}) of the resolved toric variety $X_{\widetilde\Sigma}$ and takes the form 
\begin{equation}
  \label{eq:Reqlocal}
  R_i \sim n_i\,D_i + \sum_k E_k.
\end{equation}
This relation is independent from the chosen triangulation. Since such a relation holds for every fixed point $\zf{i}{\alpha}$, we add the label $\alpha$ which denotes the different fixed sets in the $i$--direction. Furthermore, we have to sum over all fixed sets which lie in the respective fixed plane $D_{i,\alpha}$:
\begin{equation}
  \label{eq:Reqglobal}
  R_i \sim n_i\,D_{i,\alpha} + \sum_{k, \beta} E_{k\alpha\beta} \qquad \mathrm{ \;for\; all\; } \alpha \mathrm{ \; and \; all\;} i,
\end{equation}
where $n_i$ is the order of the group element that fixes the plane $D_{i,\alpha}$. The precise form of the sum over the exceptional divisors depends on the singularities involved.

In general, an orbifold of the form $T^6/G$ has local singularities of the form $\IC^m/H$, where $H$ is some subgroup of index $p = [G:H]$ in $G$. If $H$ is a strict subgroup of $G$, the above discussion applies in exactly the same way and yields relations (\ref{eq:Reqlocal}) for divisors $R'_i$ with vanishing orders $n_i'$. In the end, however, it must be taken into account that $H$ is a subgroup, which means that the relations for the $R_i'$ with the action of $H$ must be embedded into those involving the $R_i$ with the action of $G$. The $R_i'$ are related to the $R_i$ by
\begin{equation}
  \label{eq:ReqGH}
  R_i = \frac{|G|}{|H|} R_i' = p\, R_i'.
\end{equation}

When a set is fixed only under a strict subgroup $H\subset G$, its elements are mapped into each other by the generator of the normal subgroup $G/H$. Therefore, the equivalence classes of invariant divisors must be considered. They are represented by $S = \sum_{\alpha} \widetilde{S}_{\alpha}$, where $\widetilde{S}_{\alpha}$ stands for any divisor $\Dt_{i\alpha}$ or $\Et_{k\alpha\beta}$ on the cover and the sum runs over the $p$ elements of the coset $G/H$. In this case, we can add up the corresponding relations:
\begin{equation}
  \sum_\alpha R_i' \sim n_i'\sum_{\alpha} \Dt_{i\alpha} + \sum_{k,\beta} \sum_{\alpha} \Et_{k\alpha\beta}\,.
\end{equation}
The left hand side is equal to $p\, R_i' = R_i$, therefore
\begin{equation}
  \label{eq:ReqH}
  R_i \sim n_i' D_i + \sum_{k,\beta} E_{k\beta}\,,
\end{equation}
which is the same as the relation for $R_i'$. 

Something special happens if $n_i = n_j = n$ for $i \not= j$. In this
situation, there are additional divisors on the cover,
\begin{equation}
  R_{ij} = \bigcup_{k=1}^n \{z^i + \varepsilon^k z^j = \varepsilon^{k+k_0} c^{ij} \}
\end{equation}
for some integer $k_0$ and some constant $c_{ij}$, which descend to divisors on the orbifold. We have $\varepsilon^n = 1$ for even $n$, and $\varepsilon^{2n} = 1$ for odd $n$. Since the natural basis for $H^2(T^6)$ are the forms $h_{i\jbar}$ (see the previous subsection), we have to combine the various components of the $R_{ij}$ in a particular way in order to obtain divisors $R_{i\jbar}$ which are Poincar\'e dual to these forms. If we define the variables
\begin{align}
  \label{eq:zij}
  z^{ij}_{\pm} &= z^i \pm z^j, & z'_{\pm}{}^{ij} &= z^i \pm \varepsilon z^j,\\
  z^{ij}_{k} &= z^i + \varepsilon^k z^j, 
\end{align}
then 
\begin{align}
  \label{eq:Rij}
  R_{i\jbar} &= \{z^{ij}_+ + \bar{z}^{ij}_- = c^{ij} \} \cup \{z^{ij}_+ - \bar{z}^{ij}_- = c^{ij} \} \cup \{ z'_+{}^{ij} + \bar{z}'_-{}^{ij} = c^{ij} \} \cup \{ z'_+{}^{ij} - \bar{z}'_-{}^{ij} = c^{ij} \}.
\end{align}
These divisors again satisfy linear relations of the form~\eqref{eq:Reqglobal}:
\begin{align}
  \label{eq:Rijeqglobal}
  R_{i\jbar} \sim n D_{i\jbar\alpha} + \sum_{k, \beta, \gamma} E_{k\alpha\beta\gamma}.
\end{align}


\begin{exa}
  \label{sec:relexA}

  This example combines several complications: More than three
  inherited exceptional divisors, several kinds of local patches for
  the fixed points, and fixed sets which are in orbits with length
  greater than one.

  The $D$--planes are $\Dt_{1,\alpha} = \{ z^1=\zf{1}{\alpha}\}$,
  $\alpha=1,\dots,6$, $\Dt_{2,\beta} = \{z^2=\zf{2}{\beta}\}$,
  $\beta=1,...,6$, and $\Dt_{3,\gamma} = \{z^3 = \zf{3}{\gamma}\}$,
  $\gamma=1,2,3$ on the cover. From these, we define the invariant
  combinations
  \begin{align*}
    D_{1,1} &= \Dt_{1,1}, & D_{1,2} &= \Dt_{1,2} + \Dt_{1,4} +
    \Dt_{1,6}, &
    D_{1,3} &= \Dt_{1,3} + \Dt_{1,5}, \\
    D_{2,1} &= \Dt_{2,1}, & D_{2,2} &= \Dt_{2,2} + \Dt_{2,4} + \Dt_{2,6}, & D_{2,3} &= \Dt_{2,3} + \Dt_{2,5} ,   \\
    D_{3,\gamma} &= \Dt_{3,\gamma}.
  \end{align*}
  Now, we will construct the global linear
  relations~(\ref{eq:Reqglobal}).  For this, we need the local
  equivalence relations in homology, determined in the toric patches.
  For the $\IZ_{6-I}$--patches, we have (rearranged such, that only
  one $D$ appears in each relation)
  \begin{eqnarray}
    \label{eq:lineqsixI}
    0 &\sim& 6\,D_{{1}}+2\,E_{{2}}+E_{{1}}+3\,E_{{3}},\nonumber\\
    0 &\sim& 6\,D_{{2}}+2\,E_{{2}}+E_{{1}}+3\,E_{{3}},\nonumber\\
    0 &\sim& 3\,D_{{3}}+E_{{2}}+2\,E_{{1}}.
  \end{eqnarray}
  For the $\IZ_{3}$--patches, we have
  \begin{equation}\label{eq:lineqthree}
    0 \sim 3\,D_i + E,\qquad i=1,\dots,3.
  \end{equation}
  The divisor $E$ is conceptually the same as $E_2$ in the
  $\IZ_{6-I}$--patch, which also stems from the $\IZ_3$--element, thus
  we will label it as $E_2$ in the following. To embed relation
  (\ref{eq:lineqthree}) into the global relations, we must multiply it
  by two, since $\IZ_3$ has index two in $\IZ_{6-I}$.  The local
  relation for the resolved $\IZ_2$ fixed line is
  \begin{equation}
    \label{eq:Z6IrelD12}
    0  \sim 2\,D_{{1}}+\, E_{{3}},
  \end{equation}
  where the exceptional divisor obviously corresponds to $E_3$ in the
  $\IZ_{6-I}$--patch. This relation will have to be multiplied by
  three for the global case.

  The $D_{1,1}$--plane contains three equivalence classes of
  $\IZ_{6-I}$--patches, three equivalence classes of
  $\IZ_{3}$--patches, and two equivalence classes of $\IZ_2$--fixed
  lines.  The global relation is thus obtained from the local
  relations above:
  \begin{equation}
    \label{eq:exArelD11} 
    R_1 \sim6\,D_{{1,1}}+\sum_{\gamma=1}^3E_{{1,\gamma}} +2\, \sum_{\mu=1}^2 \sum_{\gamma=1}^3 E_{{2,\mu,\gamma}}+3\, \sum_{\nu=1,2} E_{{3,\nu}}.
  \end{equation}
  The divisor $D_{1,2}$ only contains two equivalence classes of
  $\IZ_2$ fixed lines:
  \begin{equation}
    R_1 \sim2\,D_{{1,2}}+\sum_{\nu=3}^6 E_{{3,\nu}}.
  \end{equation}
  Next, we look at the divisor $D_{1,3}$, which only contains $\IZ_3$
  fixed points. The local linear equivalences~(\ref{eq:lineqthree})
  together with~(\ref{eq:ReqH}) lead to
  \begin{equation}
    \label{eq:Z6Irel13}
    R_1  \sim 3\,D_{{1,3}}+\sum_{\mu=3}^5 \sum_{\gamma=1}^3 E_{{2,\mu,\gamma}}.
  \end{equation}
  The linear relations for $D_{2,\beta}$ are the same as those for
  $D_{1,\alpha}$:
  \begin{eqnarray}
    \label{eq:exarelD2}
    R_2&\sim&6\,D_{{2,1}}+\sum_{\gamma=1}^3 E_{{1,\gamma}}+2\, \sum_{\mu=1,3} \sum_{\gamma=1}^3 E_{{2,\mu,\gamma}}+3\, \sum_{\nu=1,3} E_{{3,\nu}},\nonumber\\
    R_2&\sim&2\,D_{{2,2}}+\, \sum_{\nu=2,4,5,6} E_{{3,\nu}}, \nonumber \\
    R_2&\sim&3\,D_{{2,3}}+\sum_{\mu=2,4,5} \sum_{\gamma=1}^3 E_{{2,\mu,\gamma}}.
  \end{eqnarray}
  Finally, the relations for $D_{3,\gamma}$ are again obtained
  from~(\ref{eq:lineqsixI}):
  \begin{equation}
    \label{eq:exArelD3}
    R_3 \sim3\,D_{{3,\gamma}}+2\, E_{1,\gamma} + \sum_{\mu=1}^5 E_{{2,\mu,\gamma}} \qquad \gamma=1,\dots,3.
  \end{equation}
\end{exa}


\section{The intersection ring}\label{sec:intersections}

Here, I discuss the method of calculating the intersection ring of the resolved toroidal orbifold. We proceed analogously to the construction in Section~\ref{sec:mori} for the local patches. Recall that first, the intersection numbers between three distinct divisors were determined, and then the linear relations were used to compute all the remaining intersection numbers. In the global situation we proceed in the same way. 

With the local and global linear relations worked out in the last section at our disposal, we can determine the intersection ring as follows. First we compute the intersection numbers including the $R_i$ between distinct divisors. 
Then, we make use of the schematic picture of the fixed set configuration, see Section \ref{sec:schematic}, from which we can read off which of the divisors coming from different fixed sets never intersect. With the necessary input of all intersection numbers with three different divisors, all other intersection numbers can be determined by using the global linear equivalences (\ref{eq:Reqglobal}). 

The intersections between distinct divisors $D_{i\alpha}$ and $E_{k\alpha\beta\gamma}$ are those computed in the local patch, see Section~\ref{sec:mori}. The intersections between $R_j$ and $D_{i\alpha}$ are easily obtained from their defining polynomials on the cover. The intersection number between $R_1$, $R_2$, and $R_3$ is simply the number of solutions to 
\begin{equation}
  \{\left(\widetilde{z}^1\right)^{n_1} = c_1^{n_1},\, \left(\widetilde{z}^2\right)^{n_2} = c_2^{n_2},\, \left(\widetilde{z}^3\right)^{n_3} = c_3^{n_3}\} \, ,
\end{equation}
which is $n_1n_2n_3$. Taking into account that we calculated this on the cover, we need to divide by $|G|$ in order to get the result on the orbifold. Similarly, the divisors $D_{i\alpha}$ are defined by linear equations in the $\widetilde{z}^i$, hence we set the corresponding $n_i$ to 1. Therefore,
\begin{align}
  \label{eq:R1R2R3}
  R_1R_2R_3 &= \frac{1}{|G|} n_1n_2n_3 & R_iR_jD_{k\alpha} &= \frac{1}{|G|} n_in_j & R_iD_{j\alpha}D_{k\beta} = \frac{n_i}{|G|}
\end{align}
for $i,\, j,\, k$ pairwise distinct, and all $\alpha$ and $\beta$. Furthermore, $R_i$ and $D_{i\alpha}$ never intersect by definition. The only remaining intersection numbers involving both $R_j$ and $D_{i\alpha}$ are of the form $R_jD_{i\alpha}E_{k\alpha\beta\gamma}$. They vanish if $D_{i\alpha}$ and $E_{k\alpha\beta\gamma}$ do not intersect in the local toric patch, otherwise they are 1. Finally, there are the intersections between $R_i$ and the exceptional divisors. If the exceptional divisor lies in the interior of the toric diagram or on the boundary adjacent to $D_{i\alpha}$, it cannot intersect $R_i$. Also, $R_iR_jE_{k\alpha\beta\gamma} = 0$. The above can also be seen directly from a schematic picture such as Figure \ref{fig:divisors}, combined with the toric diagrams of the local patches. 

Using this procedure it is also straightforward to compute the intersection numbers involving the divisors $R_{i\jbar}$ and $D_{i\jbar}$. From the defining polynomials in~\eqref{eq:Rij} we find that the only non--vanishing intersection numbers are
\begin{align}
  \label{eq:R1R12R21}
  R_{i\jbar}R_{j\ibar}R_k &= -\frac{1}{|G|} n_i^2n_k, & D_{i\jbar\alpha}R_{j\ibar}R_k &= -\frac{1}{|G|} n_i n_k, & R_{i\jbar}R_{j\ibar}D_{k\alpha} &= -\frac{1}{|G|} n_i^2,\notag\\
  D_{i\jbar\alpha}D_{j\ibar\beta}R_k & = -\frac{1}{|G|} n_k, &  D_{i\jbar\alpha}R_{j\ibar}D_{k\beta} & = -\frac{1}{|G|} n_i, & D_{i\jbar\alpha}D_{j\ibar\beta}D_{k\gamma} &= -\frac{1}{|G|}, \notag\\
  R_{i\jbar}R_{j\kbar}R_{k\ibar} &= \frac{1}{|G|} n_i^3, &  R_{i\jbar}R_{j\kbar}D_{k\ibar\alpha} &= \frac{1}{|G|} n_i^2, &  R_{i\jbar}D_{j\kbar\alpha}D_{k\ibar\beta} &= \frac{1}{|G|} n_i, \notag\\
  D_{i\jbar\alpha}D_{j\kbar\beta}D_{k\ibar\gamma} &= \frac{1}{|G|},
\end{align}
for $i,\, j,\, k$ pairwise distinct, and all $\alpha$, $\beta$, and $\gamma$. The negative signs come from carefully taking into account the orientation reversal due to complex conjugation. 

Using the linear relations~(\ref{eq:Reqglobal}) which take the general form 
\begin{equation}
  \sum_{a} n_s S_a = 0 \, ,
\end{equation}
we can construct a system of equations for the remaining intersection numbers involving two equal divisors $S_{aab}$ and three equal divisors $S_{aaa}$ by multiplying the linear relations by all possible products $S_bS_c$. This yields a highly overdetermined system of equations 
\begin{equation}\label{system}
\sum_a n_a S_{abc} = 0,
\end{equation} 
whose solution determines all the remaining intersection numbers. Since there are as many relations as global $D$ divisors, it is possible to eliminate the $D$s completely.

The intersection ring can also be determined without solving the system of equations (\ref{system}). All that is needed are the intersection numbers obtained from the compactified local patches and the configuration of the fixed sets. If such a patch has no exceptional divisors on the boundary of the uncompactified toric diagram, the intersection numbers of these exceptional divisors remain unchanged in the global setting. 
If the intersection number involves exceptional divisors on the boundary of the toric diagram, the local intersection number must be multiplied with the number of patches which sit on the fixed line to which the exceptional divisor belongs.


\begin{exa}
  \label{exAint}

  After the preparations of Section \ref{sec:relexA}, we are ready to
  compute the intersection ring for this example.  With $n_1=n_1=6$,
  $n_3=3$ and $|G|=6$, we obtain the following intersection numbers
  between three distinct divisors:
  \begin{align}
    \label{eq:intZ6I}
    R_1R_2R_3 &= 18, & R_1R_2D_3 &= 6, & R_1R_3D_2 &=3, & R_1D_2D_3 &= 1, \notag\\
    R_2R_3D_1 &=  3, & R_2D_1D_3 &= 1, & R_3D_1E_3 &=1, & R_3D_2E_3 &= 1, \notag\\
    D_1E_1D_3 &=  1, & D_1E_1E_2 &= 1, & D_1E_2E_3 &=1, & D_2D_3E_1 &= 1. \notag\\
    D_2E_1E_2 &= 1, & D_2E_2E_3 &= 1,
  \end{align}
  Now, we add the labels $\alpha,\beta,\gamma$ of the fixed points to
  the divisors: $D_i \to D_{i\alpha}$, $E_1 \to E_{1\gamma}$, $E_2 \to
  E_{2\alpha\beta\gamma}$, $E_3 \to E_{3\alpha}$, and set $\alpha=1,
  \beta=1, \gamma=1,2,3$.

  The global information comes from the linear relations and the
  examination of Figure~\ref{fig:ffixedi} to determine those pairs of
  divisors which never intersect. Solving the resulting overdetermined
  system of linear equations then yields the intersection ring of $X$
  in the basis $\{R_i, E_{k\alpha\beta\gamma}\}$:
  \begin{align}
    \label{eq:ringZ6Ia}
    R_1R_2R_3 &= 18, & R_3E_{3,1}^2 &= -2, & R_3E_{3,\nu}^2 &= -6, & E_{1,\gamma}^3 &= 8, \notag\\
    E_{1,\gamma}^2 E_{2,1,\gamma} &= 2, & E_{1,\gamma}E_{2,1,\gamma}^2 &= -4, & E_{2,1\gamma}^3 &= 8, & E_{2,\mu,\gamma}^3 &= 9, \notag\\
    E_{2,1,\gamma}E_{3,1}^2 &= -2, & E_{3,1}^3 &= 8,
  \end{align}
  for $\mu=2,\dots,5$, $\nu=2,\dots,6$, $\gamma=1,2,3$.
\end{exa}


\section{Divisor topologies for the compact manifold}\label{sec:divtopII}

In Section \ref{sec:ExcepTop}, the topology of the compact factors of the exceptional divisors was determined in the setting of the local non--compact patches. Here, we discuss the divisor topologies in the compact geometry of the resolved toroidal orbifolds, \ie\ in particular the topologies of the formerly non--compact $\IC$--factor of the semi--compact exceptional divisors and the topologies of the $D$--divisors about which we could not say anything in the local toric setting.

For both the exceptional divisors and the $D$--divisors, we have to distinguish two cases:
\begin{itemize}
\item[a)] The divisors belongs to a fixed set which is \emph{alone} in its equivalence class
\item[b)] The divisors belongs to a fixed set which is in an equivalence class with $p$ elements.
\end{itemize}
\subsubsection{Topologies of the exceptional divisors}

The topology of the exceptional divisors depends on the structure of the fixed point set they originate from. The following three situations can occur:
\renewcommand{\labelenumi}{E\theenumi)}
\begin{enumerate}
  \item Fixed points
  \label{item:E1}
  \item Fixed lines without fixed points
  \label{item:E2}
  \item Fixed lines with fixed points on top of them
  \label{item:E3}
\end{enumerate}
We first discuss the case a). The topology of the divisors in case~E\ref{item:E1}) has already been discussed in great detail in Section~\ref{sec:ExcepTop}. The local topology the divisors in the cases~E\ref{item:E2}) and~E\ref{item:E3}) has also been discussed in that section, and found to be (a blow--up of) $\IC\times\IP^1$. The $\IC$ factor is the local description of the $T^2/\IZ_k$ curve on which there were the $\IC^2/\IZ_m$ singularities whose resolution yielded the $\IP^1$ factor. 

For the determination of the topology of the resolved curves, it is necessary to know the topology of $T^2/\IZ_k$. This can be determined from the action of $\IZ_k$ on the respective fundamental domains. For $k=2$, there are four fixed points at 
\begin{equation}
  0, 1/2, \tau/2, \text{ and } (1+\tau)/2 
\end{equation}
for arbitrary $\tau$. The fundamental domain for the quotient can be taken to be the rhombus $[0,\tau,\tau+1/2,1/2]$ and the periodicity folds it along the line $[\tau/2,(1+\tau)/2]$. Hence, the topology of $T^2/\IZ_2$ without its singularities is that of a $\IP^1$ minus 4 points. 

\begin{figure}[h!]
\begin{center}
\includegraphics[width=.85\textwidth]{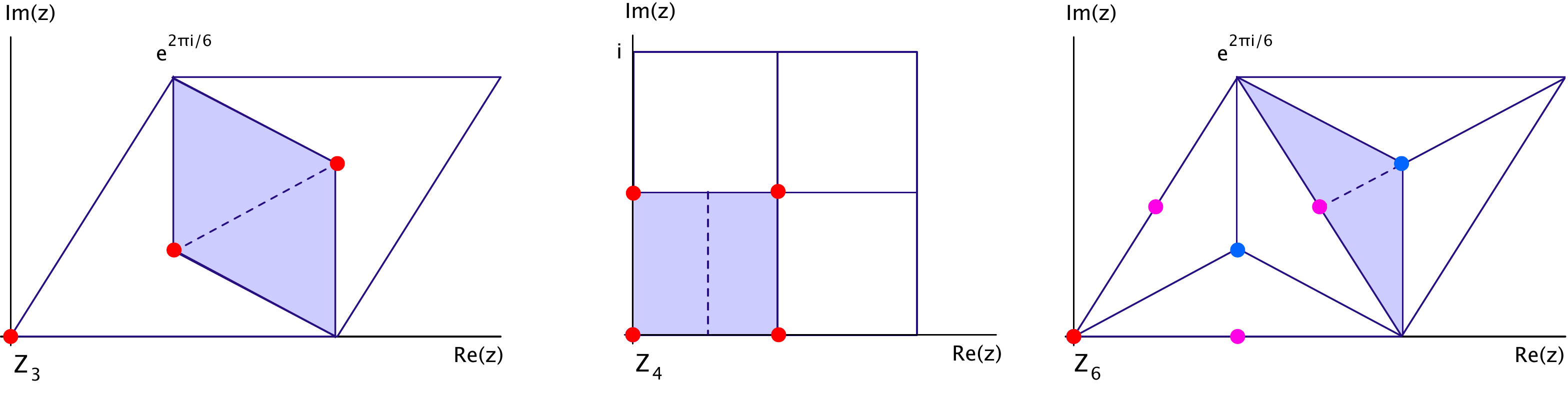}
\caption{The fundamental domains of $T^2/\IZ_k$, $k=3,4,6$. The dashed line indicates the folding.}
\label{fig:fundomains}
\end{center}
\end{figure}

For $k=3,4,6$ the value of $\tau$ is fixed to be $i, \exp(\frac{2\pi i}{3}), \exp(\frac{2\pi i}{6})$, respectively, and the fundamental domains are shown in Figure~\ref{fig:fundomains}. 

From this figure, we see that the topology of $T^2/\IZ_k$ for $k=3,4,6$ is that of a $\IP^1$ minus 3, 2, 3 points, respectively. 
\begin{itemize}
\item[E2)] There are no further fixed points, so the blow--up procedure merely glues points into this $\IP^1$. The topology of such an exceptional divisor is therefore the one of $\IF_0 = \IP^1 \times \IP^1$. 
\item[E3)] The topology further depends on the fixed points lying on these fixed lines. This depends on the choice of the root lattice for $T^6/G$, and can therefore only be discussed case by case.
\end{itemize}
The general procedure consists of looking at the corresponding toric diagram. There will always be an exceptional curve whose line ends in the point corresponding to the exceptional divisor. 
This exceptional curve meets the $\IP^1$ (minus some points) we have just discussed in the missing points and therefore, the blow--up adds in the missing points. Any further lines ending in that point of the toric diagram correspond to additional blow--ups, \ie\ additional $\IP^1$s that are glued in at the missing points. Therefore, for each fixed point lying on the fixed line and each additional line in the toric diagram there will be a blow--up of $\IF_0=\IP^1 \times \IP^1$.

In case b), \ie\ if there are $p$ elements in the equivalence class of the fixed line, the topology is quite different for the case~E\ref{item:E2}). This is because the $p$ different $T^2/\IZ_k$'s are mapped into each other by the corresponding generator in such a way that the different singular points are permuted. When the invariant combinations are constructed by summing over all representatives, the singularities disappear and we are left with a $T^2$. Hence, in the case~E\ref{item:E2}) without fixed points, the topology of $E= \sum_{\alpha=1}^k \Et_\alpha$ is $\IP^1 \times T^2$. 

\subsubsection{Topologies of the $D$--divisors}

Similarly, the topology of the divisors $D_{i\alpha}$ depends on the
structure of the fixed point sets lying in the divisor. We again treat
first case a).  Recall that the $D$--divisors are defined by
$D_{i\alpha} = \{z^i = \zf{i}{\alpha}\}$. The orbifold group $G$ acts
on these divisors by
\begin{equation}
  (z_j,z_k) \to (\varepsilon^{n_j}z_j,\varepsilon^{n_k} z_k) \text{ for } (z_j,z_k)\in D_{i\alpha} \text{ and } j\not=i\not=k \, .
\end{equation}
Since $n_j+n_k = n - n_i < n$, the resolved space will not be a
Calabi--Yau manifold anymore, but a rational surface. This happens
because for resolutions of this type of action, the canonical class
cannot be preserved. (In more mathematical terms, the resolution is
not crepant.) In order to determine the topology, we will use a
simplicial cell decomposition, remove the singular sets, glue in the
smoothening spaces, \ie\ perform the blow--ups, and use the additivity
of the Euler number. This has to be done case by case. If, in
particular, the fixed point set contains points, there will be a
blow--up for each fixed point and for each line in the toric diagram
of the fixed point which ends in the point corresponding to
$D_i$. Another possibility is to apply the techniques of toric
geometry given in Section~\ref{sec:resolve} to singularities of the
form $\IC^2/\IZ_n$ for which $n_1+n_2 \not= n$.

In case b), the basic topology again changes to $\IP^1\times T^2$.

Note that when embedding the divisor $D$ into a (Calabi--Yau) manifold $X$ in general, not all the divisor classes of $D$ are realized as classes in $X$. In the case of resolved torus orbifolds, this happens because the underlying lattice of $D$ is not necessarily a sublattice of the underlying lattice of $X$. This means that the fixed point set of $D$ as a $T^4$--orbifold can be larger than the restriction of the fixed point set of the $T^6$--orbifold to $D$. In order to determine the topology of $D$, we have to work with the larger fixed point set of $D$ as a $T^4$--orbifold. It turns out that there is always a lattice defining a $T^6$--orbifold for which all divisor classes of $D$ are also realized in $X$. In fact, we observe that the topology of all those divisors which are present in several different lattices is independent of the lattice. 

\subsubsection{Topologies of the inherited divisors}

The divisors $R_i$ contain by definition no component of the fixed point set. However, they can intersect fixed lines in points. If there are no fixed lines piercing them, the action of the orbifold group is free and their topology is that of a $T^4$. Otherwise, the intersection points have to be resolved in the same way as for the divisors $D_{i\alpha}$. In this case, the topology is always that of a K3 surface.

\subsubsection{Summary}

Table~\ref{table:sum} summarizes the basics topologies for the different divisors.
\begin{table}
\begin{center}
\begin{tabular}
{ccc}\toprule
& exceptional divisors and $D$--divisors & inherited $R$--divisors \cr
\midrule
a) & $\IP^2,\ \IF_n$& \multirow{2}{5cm}{pierced by fixed lines: $K3$, not pierced by fixed lines: $T^4$}\\
\cmidrule{1-2}
b) & $\IP^1\times T^2$&\cr
\bottomrule
\end{tabular}\end{center}
\caption{Basic divisor topologies for resolved toroidal orbifolds}
\label{table:sum}
\end{table}

In Table \ref{table:char}, we collect $\chi(\Oc_S)$, $\chi(S)$, and $K_S^2$ for the basic topologies, which are characteristic quantities of a surface $S$ and are often relevant to determine the physics of the model in question. $\chi(\Oc_S)$ is the holomorphic Euler characteristic of $S$,
\begin{equation}
  \label{eq:holoEuler}
  \chi(\Oc_S) = 1 - h^{(1,0)}(S) + h^{(2,0)}(S).
\end{equation}
$\chi(S)$ is the Euler number, and $K_S$ is the canonical divisor of $S$, 
\begin{equation}
  K_S^2 = S^2 = \ch_1(S)^2 \, .
\end{equation}
The holomorphic Euler characteristic is a birational invariant, \ie\ it does not change under blow--ups. On the other hand, blowing up a surface adds a 2--cycle to it, hence increases the Euler number $\chi(S)$ by 1.
\begin{table}
\begin{equation}
  \begin{array}{ccccc}
    \toprule
    S & \chi(S) & \chi(\Oc_S) & K_S^2 & h^{(1,0)}(S) \\
    \midrule
 \rowcolor[gray]{.95}    \IP^2 & 3 & 1 & 9 & 0\\
    \IF_n & 4 & 1 & 8 & 0\\
 \rowcolor[gray]{.95}    \IP^1 \times T^2 & 0 & 0 & 0 & 1\\
    T^4 & 0 & 0 & 0 & 2\\
 \rowcolor[gray]{.95}    \mathrm{K3} & 24 & 2 & 0 & 0 \\
    \bottomrule
  \end{array}
\end{equation}
\caption{Characteristic quantities for the basic divisor topologies}
\label{table:char}
\end{table}


\begin{exa}
  \label{sec:divsixiglobal}

  Here, we discuss the topologies of the divisors of the resolution of
  $T^6/\IZ_{6-I}$ on $G_2^2\times SU(3)$. The topology of the compact
  exceptional divisors has been determined in
  Section~\ref{sec:exsixiaaa}: $E_{1,\gamma} = \IF_4$ and
  $E_{2,1,\gamma} = \IF_2$. With the methods of toric geometry, we
  find the exceptional divisors coming from the resolution of the
  $\IZ_3$--patch, $E_{2,\mu,\gamma}$, $\mu=2,\dots,5$, to have the
  topology of a $\IP^2$. The divisor $E_{3,1}$ is of
  type~E\ref{item:E3}) and has a single representative, hence the
  basic topology is that of a $\IF_0$. There are 3 $\IZ_{6-I}$ fixed
  points on it, but there is only a single line ending in $E_3$ in the
  toric diagram of Figure~\ref{fig:fsixi}, which corresponds to the
  exceptional $\IP^1$, therefore there are no further blow--ups. The
  divisors $E_{3,\nu}$, $\nu=2,\dots,6$ are all of
  type~E\ref{item:E2}) with 3 representatives, hence their topology is
  that of $\IP^1\times T^2$.

  The topology $D_{2,1}$ is determined as follows: The fixed point set
  of the action $\frac{1}{6}(1,4)$ agrees with the restriction of the
  fixed point set of $T^6/\IZ_{6-I}$ to $D_{2,1}$. The Euler number of
  $D_{2,1}$ minus the fixed point set is
  \begin{equation}
    (0-4\cdot 0-6\cdot 1)/6=-1 \, .
  \end{equation}

  The procedure of blowing up the singularities glues in 3
  $\IP^1\times T^2$s at the $\IZ_2$ fixed lines which does not change
  the Euler number. The last fixed line is replaced by a $\IP^1 \times
  T^2$ minus 3 points, upon which there is still a free $\IZ_3$
  action. Its Euler number is therefore $(0-3)/3 = -1$. The 6 $\IZ_3$
  fixed points fall into 3 equivalence classes, furthermore we see
  from Figure~\ref{fig:frthree} that there is one line ending in
  $D_2$. Hence, each of these classes is replaced by a $\IP^1$, and
  the contribution to the Euler number is $3\cdot 2=6$. Finally, for
  the 3 $\IZ_{6-I}$ fixed points there are 2 lines ending in $D_2$ in
  the toric diagram in Figure~\ref{fig:fsixi}. At a single fixed
  point, the blow--up yields two $\IP^1$s touching in one point whose
  Euler number is $2\cdot 2 -1=3$. Adding everything up, the Euler
  number of $D_{2,1}$ is
  \begin{equation}
    \chi_{D_{2,1}}=-1 + 0 -1 + 6 + 3\cdot 3 = 13 \, ,
  \end{equation}
  which can be viewed as the result of a blow--up of $\IF_0$ in 9
  points. The same discussion as above also holds for $D_{1,1}$,
  however, there are no $\IZ_2$ fixed lines without fixed points. The
  topology of each representative of $D_{1,2}$ minus the fixed point
  set, viewed as a $T^4$ orbifold, is that of a
  \begin{equation}
    T^2 \times (T^2/\IZ_2 \setminus \{ 4\ \mathrm{pts} \}) \, .
  \end{equation}
  The representatives are permuted under the residual $\IZ_3$ action
  and the 12 points fall into 3 orbits of length 1 and 3 orbits of
  length 3. Hence, the topology of the class is still that of a $T^2
  \times (T^2/\IZ_2 \setminus \{ 4 \ \mathrm{pts} \})$. After the
  blow--up it is therefore a $\IP^1 \times T^2$. The divisor $D_{2,2}$
  has the same structure as $D_{1,2}$, therefore its topology is that
  of a $\IP^1\times T^2$. The topology of the divisors $D_{2,3}$ and
  $D_{1,3}$ is the same as the topology of $D_{i\alpha}$ in the
  $\IZ_3$ orbifold. It can be viewed as a blow--up of $\IP^2$ in 12
  points. Finally, there are the divisors $D_{3\gamma}$. The action
  $\frac{1}{6}(1,1)$ on $T^4$ has 24 fixed points, 1 of order 6, 15 of
  order 2, and 8 of order 3. The $\IZ_2$ fixed points fall into 5
  orbits of length 3 under the $\IZ_3$ element, and the $\IZ_3$ fixed
  points fall into 4 orbits of length 2 under the $\IZ_2$ element. For
  each type of fixed point there is a single line ending in $D_3$ in
  the corresponding toric diagram, therefore the fixed points are all
  replaced by a $\IP^1$. The Euler number therefore is
  \begin{equation}
    \chi_{D_{3,\gamma}}=(0-24)/6 + (1+5+4)\cdot 2 = 16 \, .
  \end{equation}
  Hence, $D_{3,\gamma}$ can be viewed as blow--up of $\IF_0$ in 12
  points.

  The divisors $R_1$ and $R_2$ do not intersect any fixed lines lines,
  therefore they simply have the topology of $T^4$. The divisor $R_3$
  has the topology of a K3. In Table~\ref{tab:TopZ6I}, we have
  summarized the topologies of all the divisors.


    \begin{table}[h!]
      \begin{center}
        $
        \begin{array}{c}
          \begin{array}{cccccccccccc}
            \toprule
            E_{1\gamma} & E_{2,1\gamma} & E_{2\mu\gamma} & E_{3,1} & E_{3,2} &           D_{1,1},D_{2,1}      & D_{1,2}         & D_{1,3}, D_{2,2}      & D_{3,\gamma} & R_1, R_2 & R_3 \\
            \midrule
            \IF_4       & \IF_2         & \IP^2          & \IF_0   & \IP^1\times T^2 &
            \Bl{9}\IF_n & \IP^1\times T^2 & \Bl{12}\IP^2  & \Bl{12}\IF_n & T^4      & \mathrm{K3} \\
            \bottomrule
          \end{array}
        \end{array}
        $
      \end{center}
      \caption{The topology of the divisors.}
      \label{tab:TopZ6I}
    \end{table}

\end{exa}


\chapter{The orientifold quotient}
\minitoc

\bigskip
Another construction the string theorist is confronted with regularly, is the orientifold quotient of some manifold $X$. We will introduce the orientifold quotient on the resolved toroidal orbifolds discussed in the previous lecture.


\section{Yet another quotient: The orientifold}

At the orbifold point, the orientifold projection is $\Omega\,I_6$, where $\Omega$ is the worldsheet orientation reversal and $I_6$ is an involution on the compactification manifold. In type IIB string theory with O3/O7--planes (instead of O5/O9), the holomorphic (3,0)--form $\Omega$ must transform as $\Omega \to -\Omega$. Therefore we choose
\begin{equation}
  \label{eq:I6}
  I_6:\ (z^1,z^2,z^3)\to (-z^1,-z^2,-z^3).
\end{equation}
Geometrically, this involution corresponds to taking a $\IZ_2$-quotient of the compactification manifold, \ie\
\begin{equation}
  B=X/I_6=(T^6/G)/I_6 \, .
\end{equation}

As long as we are at the orbifold point, all necessary information is encoded in (\ref{eq:I6}). To find the configuration of O3--planes, the fixed points under $I_6$ must be identified. On the covering space $T^6$, $I_6$ always gives rise to 64 fixed points, \ie\ 64 O3--planes. Some of them may be identified under the orbifold group $G$, such that there are less than 64 equivalence classes on the quotient. 
Each $\IZ_2$ subgroup of $G$ (if any) gives rise to a stack of O7--planes. The O7--planes are found by identifying the fixed planes under the combined action of $I_6$ and the generators $\theta_{\IZ_2}$ of the $\IZ_2$ subgroups of $G$.
A point $x$ belongs to a fixed set, if it fulfills
\begin{equation}
  \label{eq:fixO}
  I_6\,\theta_{\IZ_2}\,x=x+a,\quad a\in \Lambda,
\end{equation}
where $\Lambda$ is the torus lattice. Consequently, there are no $O7$--planes  in the prime cases, one stack \eg\ for $\IZ_{6-I}$ and three in the case of \eg\ $\IZ_2\times \IZ_6$, which contains three $\IZ_2$ subgroups. The number of O7--planes per stack depends on the fixed points in the direction perpendicular to the O--plane and therefore on the particulars of the specific torus lattice.


\section{When the patches are not invariant: $h^{(1,1)}_-\neq 0$}\label{h11minus}

Whenever $G$ contains a subgroup $H$ of odd order, some of the fixed point sets of $H$ will not be invariant under the global orientifold involution $I_6$ and will fall into orbits of length two under $I_6$. Some of these $I_6$--orbits may coincide with the $G$--orbits. In this case, no further effect arises. When $G$ contains in particular a $\IZ_2$ subgroup in each coordinate direction, all equivalence classes under $I_6$ and these subgroups coincide.
When certain fixed points or lines (which do not already form an orbit under $G$) are identified under the orientifold quotient, the second cohomology splits into an invariant and an anti--invariant part under $I_6$:
\begin{equation}
  H^{1,1}(X)=H^{1,1}_{+}(X)\oplus H^{1,1}_{-}(X) \, .
\end{equation}
The geometry is effectively reduced by the quotient and the moduli associated to the exceptional divisors of the anti--invariant patches are consequently no longer geometric moduli. They take the form~\cite{Grimm:2005fa}
\begin{equation}
  \label{eq:newmoduli}
  G^a=C^a_2+S\,B^a_2.
\end{equation}


\begin{exb}
  To determine the value of $h^{(1,1)}_{-}$ for this example, we must
  examine the configuration of fixed sets given in
  Table~\ref{fssixiia} and Figure~\ref{ffixsixiiaa}.a and the resolution
  of the local patch, see Figure \ref{ffixsixiiaa}.b, and determine
  the conjugacy classes of the fixed sets under the global involution
  $I_6:\,z^i\to -z^i$.
  \begin{table}[h!]\begin{center}
      \begin{tabular}{cccc}
        \toprule
        \ Group el.& Order & Fixed Set& Conj. Classes \cr
        \midrule
        \rowcolor[gray]{.95}$\theta$&6&12\ {\rm fixed\ points} &\ 12\cr
        $\theta^2$&3      &3\ {\rm fixed\ lines} &\ 3\cr
        \rowcolor[gray]{.95}$\theta^3$&2      &4\ {\rm fixed\ lines} &\ 4\cr
        \bottomrule
      \end{tabular}
      \caption{Fixed point set for $\IZ_{6-II}$--orbifold on
        $SU(2)\times SU(6)$.}\label{fssixiia}
    \end{center}\end{table}
\begin{figure}
  \centering
\subfigure[Toric diagram of the resolution of $\IC^3/\IZ_{6-II}$]{\includegraphics[width=.45\textwidth]{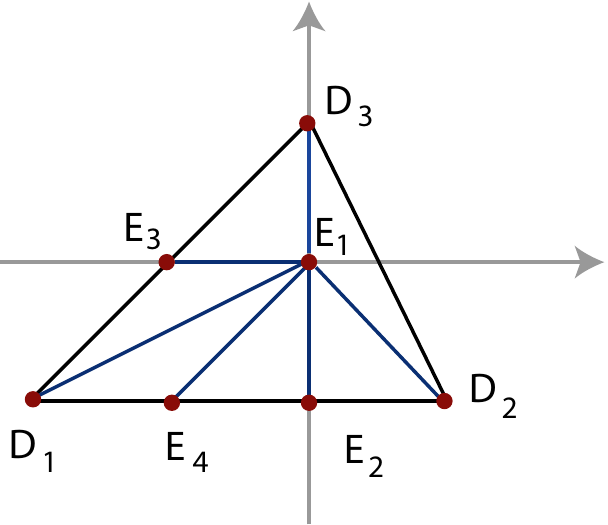}} \hfill
  \subfigure[Schematic picture of the fixed set configuration of
        $\IZ_{6-II}$ on $SU(2)\times SU(6)$]{\includegraphics[width=.45\textwidth]{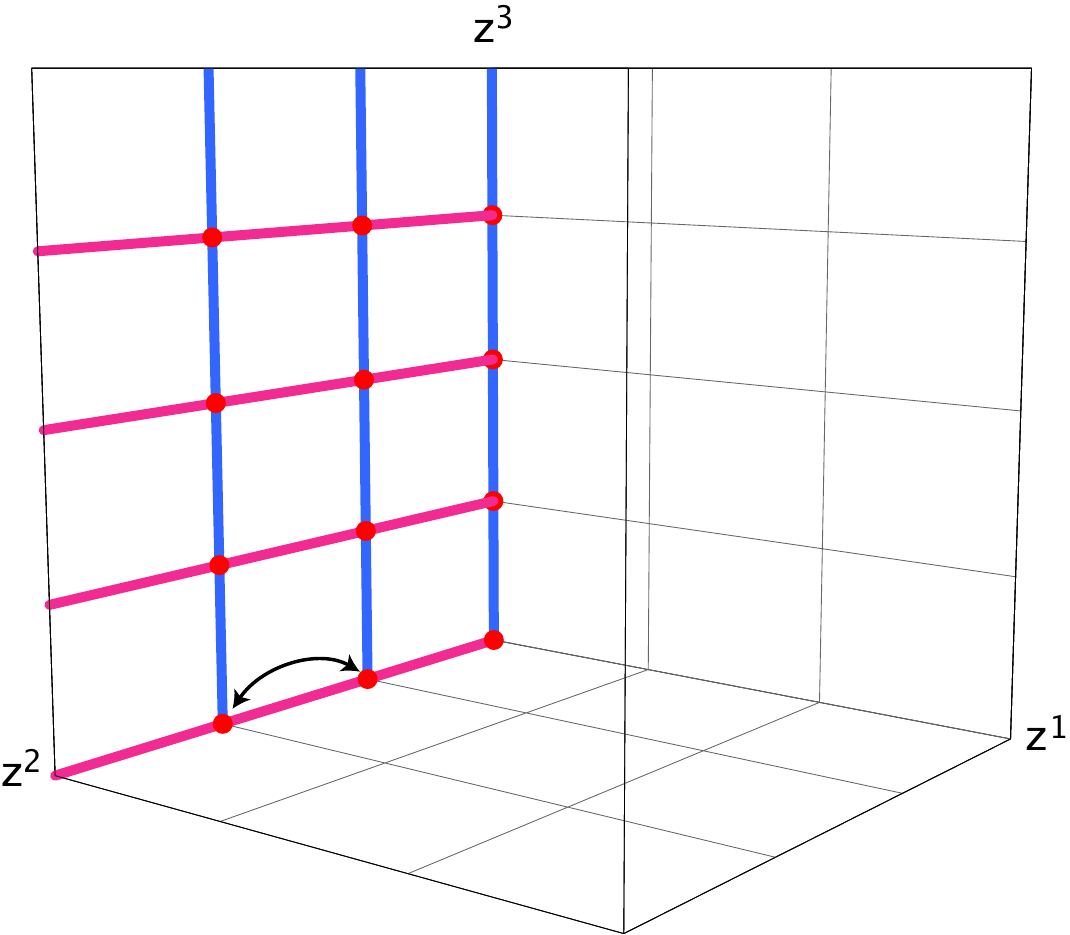}}\label{ffixsixiiaa} 
   \caption{Resolution of $\IC^3/\IZ_{6-II}$ and fixed set configuration of
        $\IZ_{6-II}$ on $SU(2)\times SU(6)$}
\end{figure}

  The fixed sets located at $z^2=0$ are invariant under $I_6$, those
  located at $z^2=1/3$ are mapped to $z^2=2/3$ and vice
  versa. Clearly, this is an example with $h^{(1,1)}_{-}\neq 0$. The
  divisors $E_{1,\beta\gamma},\, E_{2,\beta}$ and $E_{4,\beta}$ for
  $\beta=2,3$ are concerned here. Out of these twelve divisors, six
  invariant combinations can be formed:
  \begin{equation}
  E_{1,inv,\gamma}=\frac{1}{2}(E_{1,2\gamma}+E_{1,3\gamma}),\
  E_{2,inv}=\frac{1} {2}(E_{2,2}+E_{2,3})\ \text{ and }\
  E_{4,inv}=\frac{1}{2}(E_{4,2}+E_{4,3}).
  \end{equation} 
  With a minus sign instead
  of a plus sign, the combinations are anti--invariant, therefore
  $h^{(1,1)}_{-}=6$.
\end{exb}

\section[The local orientifold involution]{The local orientifold involution on the resolved patches}

Now we want to discuss the orientifold action for the smooth Calabi--Yau manifolds $X$ resulting from the resolved torus orbifolds. For such a manifold $X$, we will denote its orientifold quotient $X/I_6$ by $B$ and the orientifold projection by $\pi:X \to B$. Away from the location of the resolved singularities, the orientifold involution retains the form~(\ref{eq:I6}). As explained above, the orbifold fixed points fall into two classes:
\renewcommand{\labelenumi}{O\theenumi)}
\begin{enumerate}
  \item The fixed point is invariant under $I_6$, \ie\ its exceptional divisors are in $h^{1,1}_{+}$.
  \label{item:O1}
  \item The fixed point lies in an orbit of length two under $I_6$, \ie\ is mapped to another fixed point. The invariant combinations of the corresponding exceptional divisors contribute to $h^{1,1}_{+}$, while the remaining linear combinations contribute to $h^{1,1}_{-}$. 
  \label{item:O2}
\end{enumerate}
\renewcommand{\labelenumi}{\arabic{\theenumi}}
The fixed points of class~O\ref{item:O1}) locally feel the involution: Let $\zf{}{\alpha}$ denote some fixed point. Since $\zf{}{\alpha}$ is invariant under~(\ref{eq:I6}),
\begin{equation}
  \label{eq:onei}
  (\zf{i}{\alpha}+\Delta z^i) \to (\zf{i}{\alpha}-\Delta z^i).
\end{equation}
In local coordinates centered around $\zf{}{\alpha}$, $I_6$ therefore acts as 
\begin{equation}
  \label{eq:locali}
  (z^1, z^2, z^3) \to (-z^1,-z^2, -z^3).
\end{equation}
In case~O\ref{item:O2}), the point $\zf{}{\alpha}$ is not fixed, but gets mapped to a different fixed point $\zf{}{\beta}$. So locally,
\begin{equation}
  \label{eq:oneii}
  (\zf{i}{\alpha}+\Delta z^i) \to (\zf{i}{\beta}-\Delta z^i).
\end{equation}
In the quotient, $\zf{}{\alpha}$ and $\zf{}{\beta}$ are identified, \ie\ correspond the the same point.
In local coordinates centered around this point, $I_6$ therefore acts again as $z^i\to -z^i$, see (\ref{eq:locali}).

For the fixed lines, we apply the same prescription. The involution on fixed lines with fixed points on them is constrained by the involution on the fixed points.

What happens in the local patches after the singularities were resolved? 
A local involution $\cal I$ has to be defined in terms of the local coordinates, such that it agrees with the restriction of the global involution $I_6$ on $X$. Therefore, we require that $\cal I$ maps $z_i$ to $-z_i$. In addition to the three coordinates $z_i$ inherited from $\IC^3$, there are now also the new coordinates $y_k$ corresponding to the exceptional divisors $E_k$. For the choice of the action of $\cal I$ on the $y_k$ of an individual patch, there is some freedom. 

For simplicity we restrict the orientifold actions to be multiplications by $-1$ only. We do not take into account transpositions of coordinates or shifts by half a lattice vector. The latter have been considered in the context of toric Calabi--Yau hypersurfaces in~\cite{Berglund:1998va}. The allowed transpositions can be determined from the toric diagram of the local patch by requiring that the adjacencies of the diagram be preserved.

The only requirements ${\cal I}$ must fulfill are compatibility with the ${\IC}^*$--action of the toric variety, \ie\ 
\begin{equation}
  \label{eq:Caction}
  (-z_1,-z_2,-z_3,(-1)^{\sigma_1}y_1,\dots,(-1)^{\sigma_n}y_n) = (\prod_{a=1}^r \lambda_1^{l_1^{(a)}} z_1,\dots,\prod_{a=1}^r \lambda_n^{l_n^{(a)}} y_n)
\end{equation}
where $l_i^{(a)}$ encode the linear relations~\eqref{eq:linrels} of the toric patch, and that subsets of the set of solutions to~\eqref{eq:Caction} must not be mapped to the excluded set of the toric variety and vice versa.

The fixed point set under the combined action of ${\cal I}$ and the scaling action of the toric variety gives the configuration of O3-- and O7--planes in the local patches. Care must be taken that only these solutions which do not lie in the excluded set are considered. We also exclude solutions which do not lead to solutions of the right dimension, \ie\ do not lead to O3/O7--planes.

On an individual patch, we can in principle choose any of the possible involutions on the local coordinates. In the global model however, the resulting solutions of the individual patches must be compatible with each other. While O7--planes on the exceptional divisors in the interior of the toric diagram are not seen by the other patches, O7--plane solutions which lie on the $D$--planes or on the exceptional divisors on a fixed line must be reproduced by all patches which lie in the same plane, respectively on the same fixed line. This is of course also true for different types of patches which lie in the same plane. 

It is in principle possible for examples with many interior points of the toric diagram to choose different orientifold involutions on the different patches which lead to solutions that are consistent with each other. We choose the same involution on all patches, which for simple examples such as $\IZ_4$ or the $\IZ_6$ orbifolds is the only consistent possibility.

The solutions for the fixed sets under the combined action of ${\cal I}$ and the scaling action give also conditions to the $\lambda_i$ appearing in the scaling actions, they are set to $\pm 1$.
The O--plane solutions of the full patch descend to solutions on the restriction to the fixed lines on which the patch lies. For the restriction, we set the $\lambda_i$ which not corresponding to the Mori generators of the fixed line to $\pm1$ in accordance with the values of the $\lambda_i$ of the solution for the whole patch which lies on this fixed line.

A further global consistency requirement comes from the observation that the orientifold action commutes with the singularity resolution. A choice of the orientifold action on the resolved torus orbifold must therefore reproduce the orientifold action on the orbifold and yield the same fixed point set in the blow--down limit.

Given a consistent global orientifold action it might still happen that the model does not exist. This is the case if the tadpoles cannot be cancelled.


\begin{exb}
  \label{sec:Osixiiloc}

  On the homogeneous coordinates $y_k$, several different local
  actions are possible. We give the eight possible actions which only
  involve sending coordinates to their negatives:
  \begin{align}
    \label{eq:invlocal}
    (1)\quad\Ic:\,(z,y) &\to (-z_1,-z_2,-z_3,y_1,y_2,y_3,y_4) \notag\\
    (2)\quad \Ic:\,(z,y) &\to (-z_1,-z_2,-z_3,y_1,y_2,-y_3,-y_4) \notag\\
    (3)\quad\Ic:\,(z,y) &\to (-z_1,-z_2,-z_3,y_1,-y_2,y_3,-y_4) \notag\\
    (4)\quad\Ic:\,(z,y) &\to (-z_1,-z_2,-z_3,y_1,-y_2,-y_3,y_4) \notag\\
    (5)\quad\Ic:\,(z,y) &\to (-z_1,-z_2,-z_3,-y_1,y_2,y_3,-y_4) \notag\\
    (6)\quad\Ic:\,(z,y) &\to (-z_1,-z_2,-z_3,-y_1,y_2,-y_3,y_4) \notag\\
    (7)\quad \Ic:\,(z,y) &\to (-z_1,-z_2,-z_3,-y_1,-y_2,y_3,y_4) \notag\\
    (8)\quad \Ic:\,(z,y) &\to (-z_1,-z_2,-z_3,-y_1,-y_2,-y_3,-y_4)
  \end{align}
  In the orbifold limit, (\ref{eq:invlocal}) reduces to $I_6$.  Note
  that the eight possible involutions only lead to four distinct fixed
  sets (but to different values for the $\lambda_i$).

  We focus for the moment on the third possibility.  With the scaling
  action
  \begin{equation}\label{rescalessixiia}{(z_1,\,z_2,\,z_3,\,y_1,\,y_2,\,y_3,\,y_4)
      \to
      (\frac{\lambda_1\lambda_3}{\lambda_4}\,z_1,\,\lambda_2\,z_2,\,\lambda_3\,z_3,\,
      {1\over\lambda_4}\,y_1,\,\frac{\lambda_1}{\lambda_2^2}\,y_2,\,{\lambda_4\over
        \lambda_3^2}\,y_3, {\lambda_2\lambda_4\over \lambda_1^2}\,y_4)
    }\end{equation}
  we get the solutions
  \begin{itemize}
  \item[(i).] $z_1=0,\ \ \lambda_1=\lambda_2=\lambda_3=-1,\
    \lambda_4=1,$
  \item[(ii).] $z_3=0,\ \ \lambda_1=\lambda_2=-1,\
    \lambda_3=\lambda_4=1,$
  \item[(iii).] $y^2=0,\ \ \lambda_1=\lambda_4=1,\
    \lambda_2=\lambda_3=-1.$
  \end{itemize}
  This corresponds to an O7--plane wrapped on $D_{1}$, one on each of
  the four $D_{3,\gamma}$ and one wrapped on each of the two invariant
  $E_{2,\beta}$. No $O3$--plane solutions occur.  $\lambda_1$ and
  $\lambda_2$ correspond to the two Mori generators of the
  $\IZ_3$--fixed line. We restrict to it by setting $\lambda_3=-1,\
  \lambda_4=1$ in accordance with solution (i) and (ii) which are seen
  by this fixed line. The scaling action thus becomes
  \begin{equation}\label{rescalessixiizthree}{(z_1,\,z_2,\,z_3,\,y^1,\,y^2,\,y^3,\,y^4)
      \to
      (-\lambda_1\,z_1,\,\lambda_2\,z_2,-z_3,y^1,\,\frac{\lambda_1}{\lambda_2^2}\,y^2,
      y^3, {\lambda_2\over \lambda_1^2}\,y^4).  }\end{equation}
  $y^1$ and $y^3$ do not appear in the fixed line, and the restriction
  makes sense only directly at the fixed point, \ie\ for $z_3=0$. With
  this scaling action and the involution (3), we again reproduce the
  solutions (i) and (ii).  $\lambda_3$ corresponds to the Mori
  generator of the $\IZ_2$ fixed line. We restrict to it by setting
  $\lambda_1=\lambda_2=-1,\ \lambda_4=1$.  The scaling action becomes
  \begin{equation}\label{rescalessixiiaztwo}
    (z_1,\,z_2,\,z_3,\,y^1,\,y^2,\,y^3,\,y^4) \to (-\lambda_3\,z_1,-\,z_2,\,\lambda_3\,z_3,\,y^1,\,-y^2,\,{1\over \lambda_3^2}\,y^3, -\,y^4),
  \end{equation}
  which together with the involution (3) again reproduces the
  solutions (i) and (iii).  Global consistency is ensured since we
  only have one kind of patch on which we choose the same involution
  for all patches.
\end{exb}

\section{The intersection ring}
\label{sec:Oring}

The intersection ring of the orientifold can be determined as follows. The basis is the relation between the divisors on the Calabi--Yau manifold $X$ and the divisors on the orientifold $B$. The first observation is that the integral on $B$ is half the integral on $X$: 
\begin{equation}
  \label{eq:Ointegral}
  \int_B \widehat{S}_a \wedge \widehat{S}_b \wedge \widehat{S}_c = \frac{1}{2} \int_X S_a \wedge S_b \wedge S_c ,
\end{equation}
where the hat denotes the corresponding divisor on $B$. The second observation is that for a divisor $S_a$ on $X$ which is not fixed under $I_6$ we have $S_a = \pi^*\widehat{S}_a$. If, however, $S_a$ is fixed by $I_6$, we have to take $S_a=\frac{1}{2}\pi^*\widehat{S}_a$ because the volume of $S_a$ in $X$ is the same as the volume of $\widehat{S}_a$ on $B$. Applying these rules to the intersection ring obtained in Section~\ref{sec:intersections} immediately yields the intersection ring of $B$:  triple intersection numbers between divisors which are not fixed under the orientifold involution become halved. If one of the divisors is fixed, the intersection numbers on the orientifold are the same as on the Calabi--Yau. If two (three) of the divisors are fixed, the intersection numbers on the orientifold must be multiplied by a factor of two (four).


\begin{exb}
  \label{sec:sixiiOintersections}

  The global linear relations for the Calabi--Yau manifold are:
  \begin{eqnarray}\label{eq:globalrelsixiiaO}
    R_1&\sim&6\,D_{{1}}+3\,\sum_{\gamma=1}^4 E_{{3,\gamma}}+\sum_{\beta,\gamma}E_{{1,\beta\gamma}}+\sum_{\beta=1}^3[\,2\,E_{{2,\beta}}+4\,E_{{4,\beta}}],\cr
    R_2&\sim&3\,D_{{2,\beta}}+\sum_{\gamma=1}^4E_{{1,\beta\gamma}}+2\,E_{{2,\beta}}+E_{{4,\beta}},\cr
    R_3&\sim&2\,D_{{3,\gamma}}+\sum_{\beta=1}^3E_{{1,\beta\gamma}}+E_{{3,\gamma}}.
  \end{eqnarray}
  After the orientifold involution, they become
  \begin{eqnarray}\label{eq:globalrelsixiiaOO}
    R_1&\sim&3\,D_{{1}}+3\,\sum_{\gamma=1}^4 E_{{3,\gamma}}+\sum_{\beta,\gamma}E_{{1,\beta\gamma}}+\sum_{\beta=1}^2[\,E_{{2,\beta}}+4\,E_{{4,\beta}}],\cr
    R_2&\sim&3\,D_{{2,\beta}}+\sum_{\gamma=1}^4E_{{1,\beta\gamma}}+E_{{2,\beta}}+E_{{4,\beta}},\cr
    R_3&\sim&D_{{3,\gamma}}+\sum_{\beta=1}^2E_{{1,\beta\gamma}}+E_{{3,\gamma}}.
  \end{eqnarray}

  The intersection numbers of the Calabi--Yau are
  \begin{align}\label{iz6ii}
    R_1R_2R_3&=6,& R_3E_{2,\beta}E_{4,\beta}&=1,&
    E_{1,\beta\gamma}E_{2,\beta}E_{4,\beta}&=1,\cr
    R_2E_{3,\gamma}^2&=-2,& R_3E_{2,\beta}^2&=-2,&
    R_3E_{4,\beta}^2&=-2,\cr E_{1,\beta\gamma}^3&=6,&
    E_{2,\beta}^3&=8,& E_{3,\gamma}^3&=8,\cr E_{4,\beta}^3&=8,
    &E_{1,\beta\gamma}E_{2,\beta}^2&=-2,&
    E_{1,\beta\gamma}E_{3,\gamma}^2&=-2, \cr
    E_{1,\beta\gamma}E_{4,\beta}^2&=-2,
    &E_{2,\beta}^2E_{4,\beta}&=-2.& &\ &
  \end{align}
  Intersection numbers which contain no factor of $E_{2,\beta}$ are
  halved for the orientifold. If the intersection number contains one
  factor of $E_{2,\beta}$, it remains the same. If two (three) factors
  $E_{2,\beta}$ are present, the number on the Calabi--Yau is
  multiplied by a factor of two (four). This leads to the following
  modified triple intersection numbers:
  \begin{align}\label{iz6iiO}
    R_1R_2R_3&=3,& R_3E_{2,\beta}E_{4,\beta}&=1,&
    E_{1,\beta\gamma}E_{2,\beta}E_{4,\beta}&=1,\cr
    R_2E_{3,\gamma}^2&=-1,& R_3E_{2,\beta}^2&=-4,&
    R_3E_{4,\beta}^2&=-1,\cr E_{1,\beta\gamma}^3&=3,&
    E_{2,\beta}^3&=32,& E_{3,\gamma}^3&=4,\cr E_{4,\beta}^3&=4,
    &E_{1,\beta\gamma}E_{2,\beta}^2&=-4,&
    E_{1,\beta\gamma}E_{3,\gamma}^2&=-1, \cr
    E_{1,\beta\gamma}E_{4,\beta}^2&=-1,
    &E_{2,\beta}^2E_{4,\beta}&=-4.& &\ &
  \end{align}
\end{exb}

\section{Literature}

The methods described in Lectures 3 and 4 were pioneered in~\cite{Denef:2005mm} and later generalized in~\cite{Lust:2006zh}. An extended description can be found in~\cite{Reffert:2006du}.

\bibliographystyle{JHEP}
\bibliography{LectureReferences}


\subsection*{Acknowledgements}

I would like to thank Domenico Orlando for comments on the lectures and the manuscript, as well as Robbert Dijkgraaf for general advice. Furthermore, I would like to thank Emanuel Scheidegger for collaboration on the material covered in Lectures 3 and 4. 

Moreover, I would like to thank the organizers of the Workshop on String and M--Theory Approaches to Particle Physics and Astronomy for the giving me the possibility of teaching this lecture series, and the Galileo Galilei Institute for Theoretical Physics for hospitality, as well as INFN for partial support during the completion of this manuscript.

S.R. is supported by the EC's Marie Curie Research Training Network under the contract \textsc{MRTN}-\textsc{CT}-2004-512194 "Superstrings".


\end{document}